\documentclass[11pt]{article}

\usepackage[T1]{fontenc}
\usepackage{lmodern}
\usepackage[margin=25mm]{geometry}
\usepackage{setspace}
\usepackage{amsmath,amssymb}
\usepackage{graphicx}
\graphicspath{{figures/}}
\usepackage{booktabs}
\usepackage{longtable}
\usepackage{array}
\usepackage{xcolor}
\usepackage[authoryear,round]{natbib}
\usepackage{lineno}
\usepackage[hidelinks]{hyperref}
\usepackage{xurl} 
\hypersetup{pdftitle={The long-term divergence exponent and the temporal structure of step-to-step perturbations: a simulation study in two passive walking models},
  pdfauthor={Philippe Terrier}}

\newcommand{\lS}{\lambda_{\mathrm{S}}}
\newcommand{\lL}{\lambda_{\mathrm{L}}}
\newcommand{\sref}{\sigma_{\mathrm{ref}}}
\newcommand{\smax}{\sigma_{\max}}
\newcommand{\appref}[1]{Appendix~\ref{#1}}

\newcommand{\FigDesign}{%
\begin{figure}[tbp]
\centering
\includegraphics[width=\textwidth]{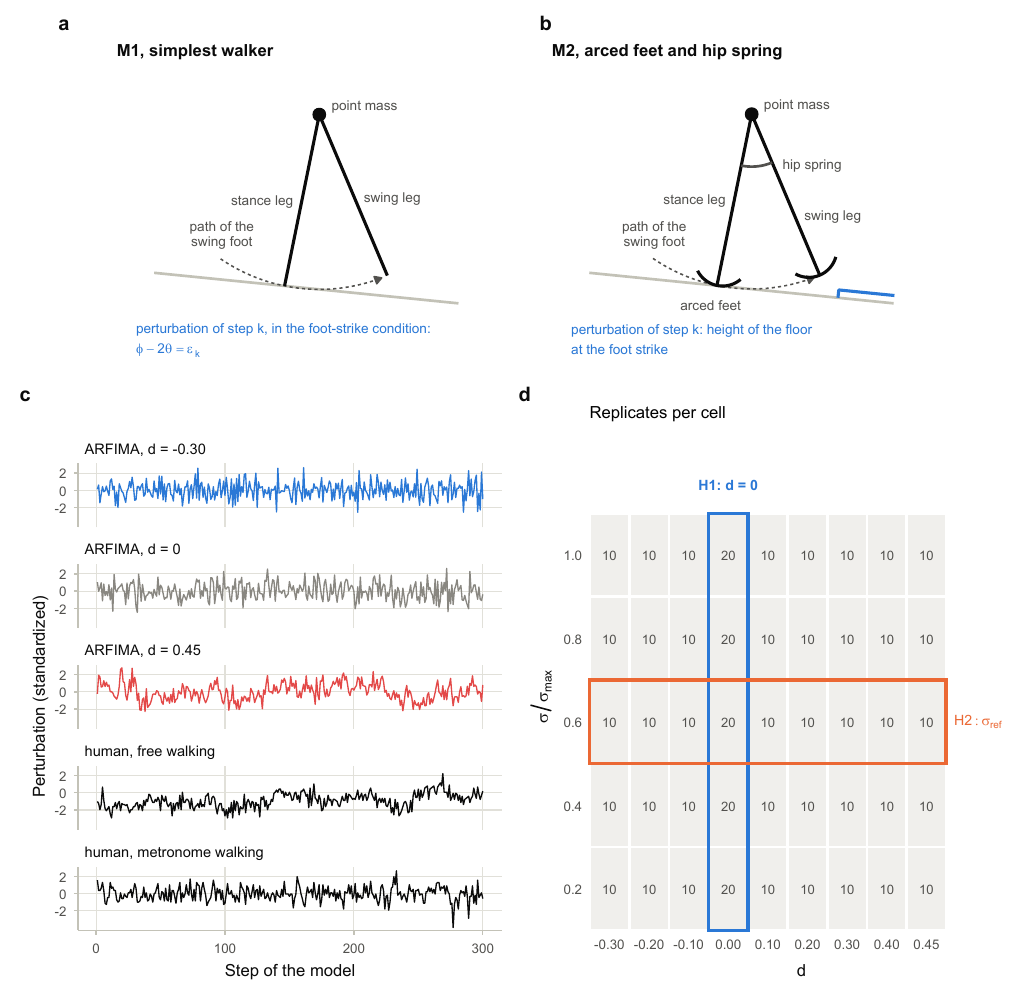}
\caption{Design of the study. (a) M1, the simplest walker, and (b) M2, the walker with arced feet and a hip spring, with the point at which the perturbation of each step enters; schematic drawings. The dashed arc is the path of the swing foot about the hip. (c) Examples of perturbation series, standardized: ARFIMA(0,\allowbreak$d$,\allowbreak0) series at three levels of $d$, and the first 300 strides of one free and one metronome series of the human database. (d) Crossed design: five levels of $\sigma$ by nine levels of $d$, with the number of replicates per cell. The frames mark the cells of the hypothesis on amplitude (H1, at $d = 0$) and of the hypothesis on temporal structure (H2, at the reference amplitude $\sref$).}
\label{fig:design}
\end{figure}}

\newcommand{\FigEstimation}{%
\begin{figure}[tbp]
\centering
\includegraphics[width=\textwidth]{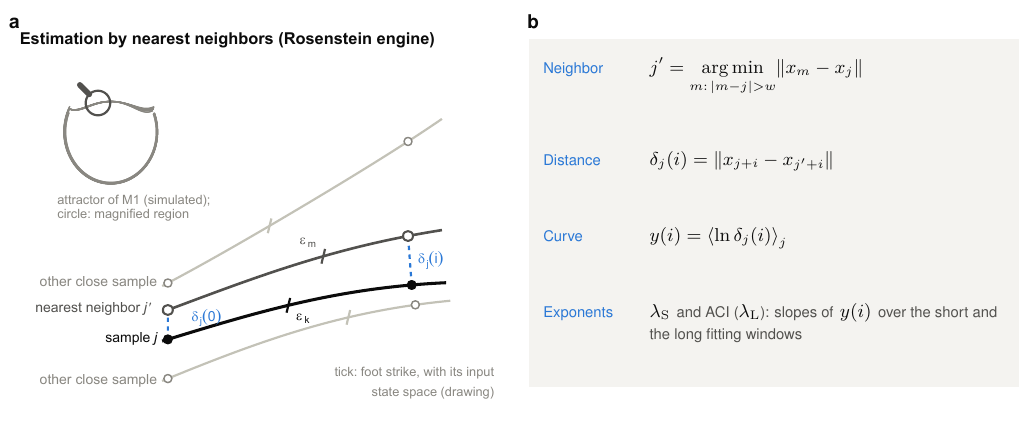}
\caption{Estimation of the divergence exponents by nearest neighbors. (a) Schematic drawing. The Rosenstein engine pairs each sample $j$ with its nearest neighbor $j'$ outside an exclusion window and follows their distance $\delta_j(i)$. Other close samples of the series (light gray) separate from the trajectory of $j$ in other ways. The inset shows the attractor of the simplest walker (M1): 20 strides of one simulated run of the reference cell, projected on the first two principal components of the state variables. Its circle marks the kind of region that the drawing magnifies. (b) Equations of the estimation; $x_j$ is the state at sample $j$, and $w$ is the exclusion window.}
\label{fig:estimation}
\end{figure}}

\newcommand{\FigCurves}{%
\begin{figure}[tbp]
\centering
\includegraphics[width=\textwidth]{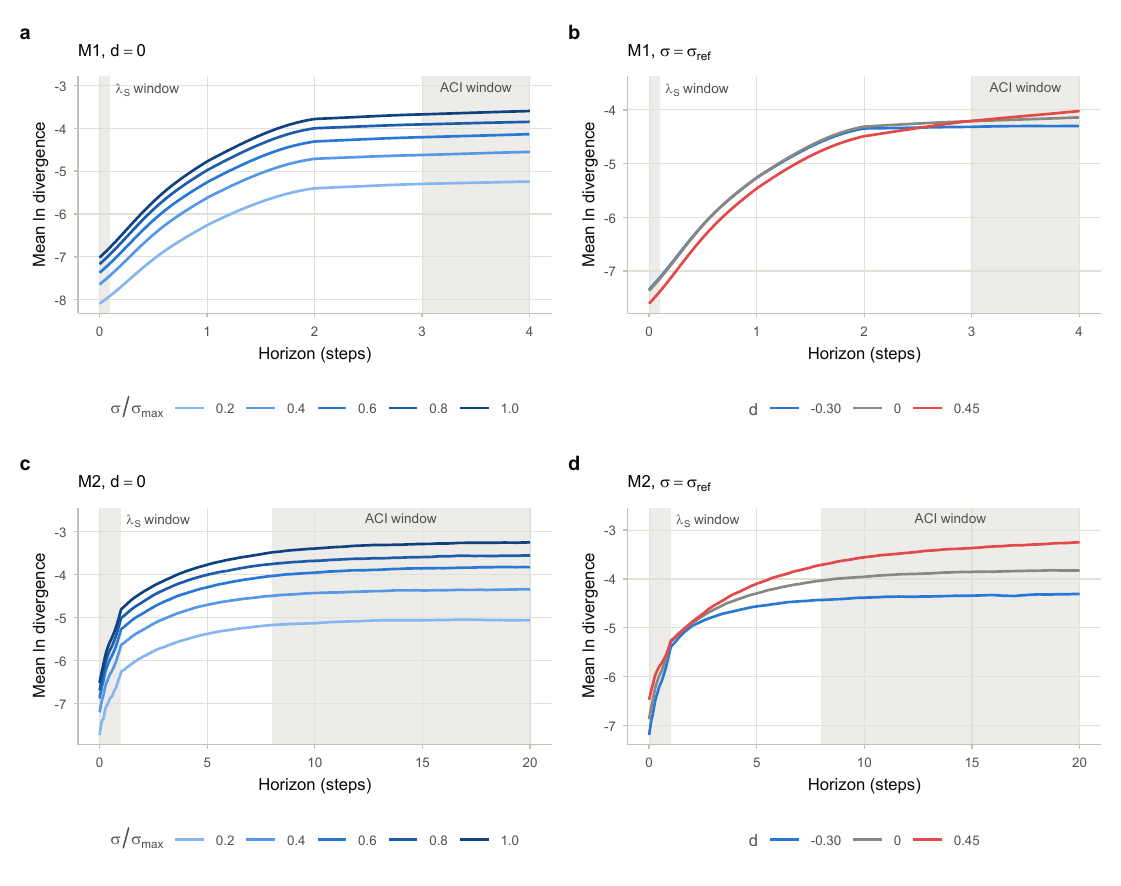}
\caption{Mean divergence curves of the simplest walker, M1 (a, b), and of the walker with arced feet and a hip spring, M2 (c, d). The curves are shown at $d = 0$ for the five levels of the amplitude $\sigma$ (a, c) and at the reference amplitude $\sref$ for three levels of $d$ (b, d). Shaded bands: fitting windows of $\lS$ and of the ACI (Table~\ref{tab:settings}). Horizon in steps.}
\label{fig:curves}
\end{figure}}

\newcommand{\FigGrid}{%
\begin{figure}[tbp]
\centering
\includegraphics[width=\textwidth]{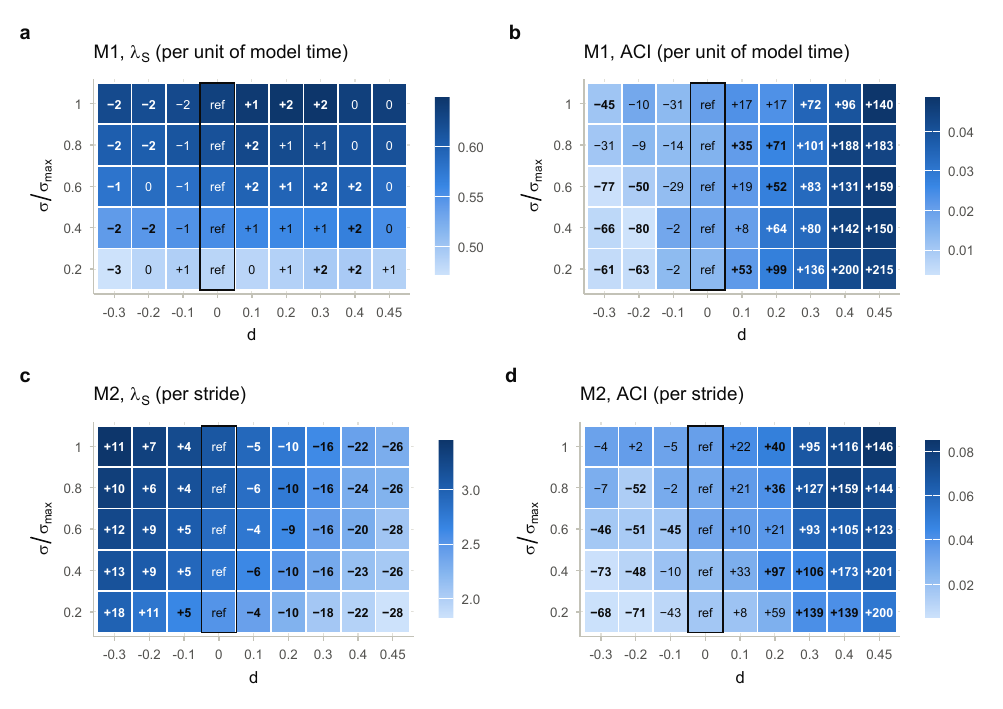}
\caption{Condition means of $\lS$ and of the ACI over the crossed design, in the simplest walker, M1 (a, b), and in the walker with arced feet and a hip spring, M2 (c, d). Each tile is one cell of the design, arranged as in Figure~\ref{fig:design}d: five levels of $\sigma/\smax$ by nine levels of $d$. Each panel has its own color scale. The number in a tile is the difference of its condition mean from the condition mean at $d = 0$ of the same amplitude (framed column, ref), in percent. Bold: the 95\% bootstrap interval of this difference excludes zero (Section~\ref{sec:exploratory}). These percent differences are descriptive.}
\label{fig:grid}
\end{figure}}

\newcommand{\FigHuman}{%
\begin{figure}[tbp]
\centering
\includegraphics[width=\textwidth]{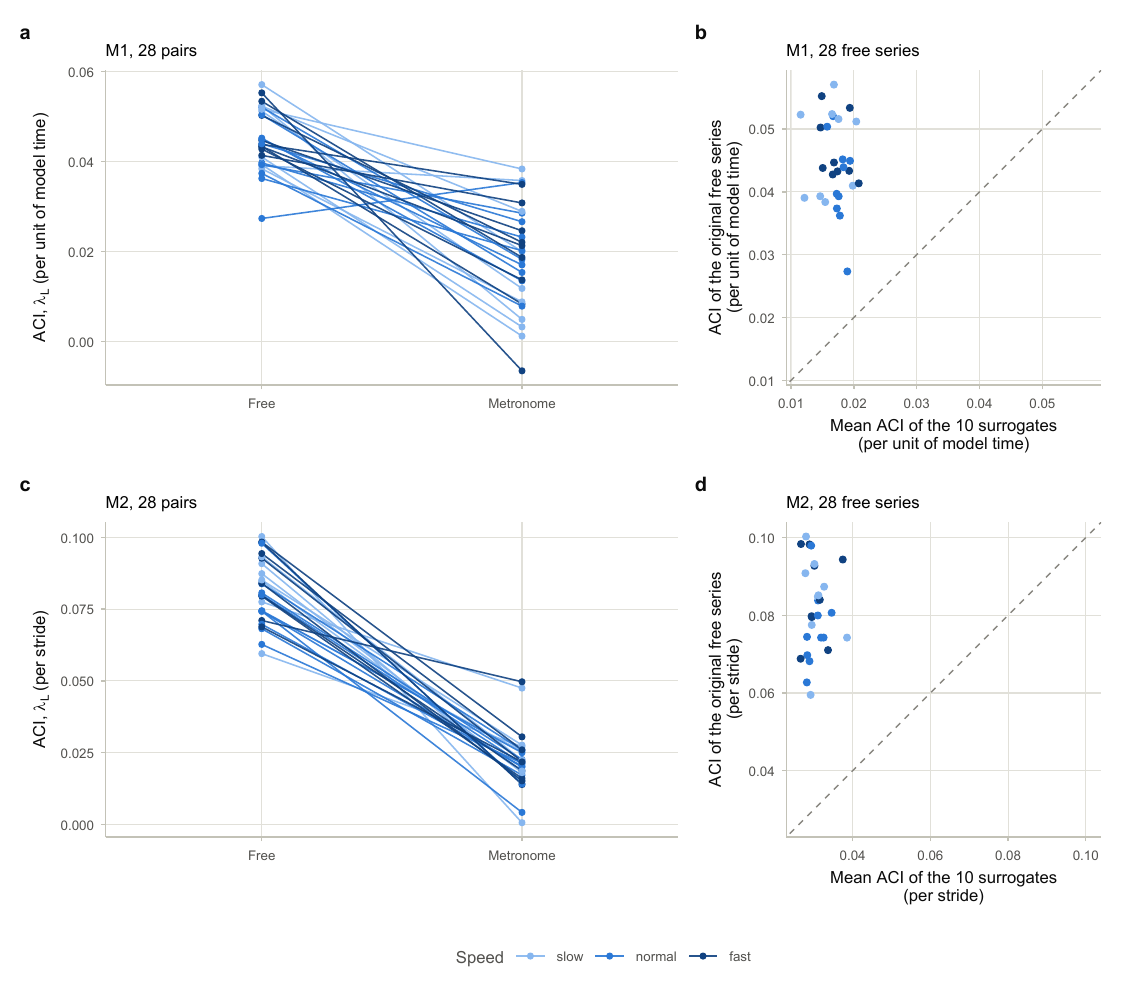}
\caption{ACI with the human stride series, in the simplest walker, M1 (a, b), and in the walker with arced feet and a hip spring, M2 (c, d). (a, c) Free and metronome series of each participant and speed, joined by a line (prediction S1). (b, d) ACI of each original free series against the mean ACI of its 10 surrogates (prediction S2); dashed line: identity.}
\label{fig:human}
\end{figure}}

\newcommand{\FigDissociation}{%
\begin{figure}[tbp]
\centering
\includegraphics[width=\textwidth]{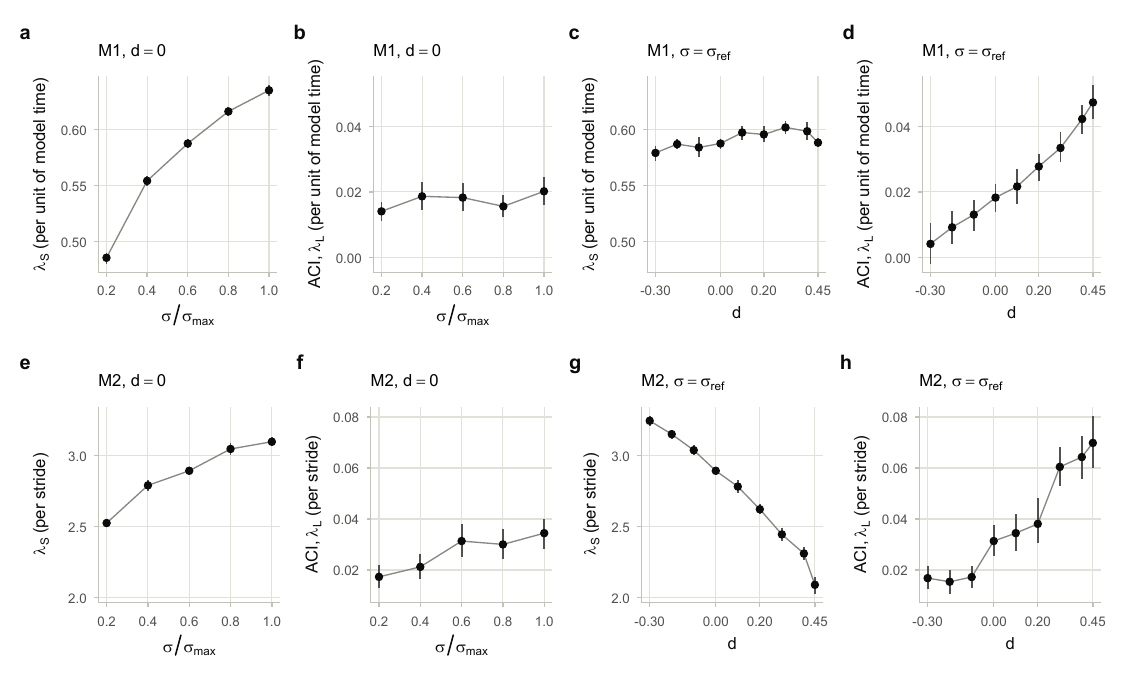}
\caption{Condition means of $\lS$ and of the ACI in the simplest walker, M1 (a to d), and in the walker with arced feet and a hip spring, M2 (e to h). The means are plotted against $\sigma/\smax$ at $d = 0$ (cells of H1, the hypothesis on amplitude) and against $d$ at the reference amplitude $\sref$ (cells of H2, the hypothesis on temporal structure). Bars: 95\% percentile bootstrap intervals (Section~\ref{sec:statistics}). Within a model, the two panels of one exponent share their vertical scale.}
\label{fig:dissociation}
\end{figure}}

\title{The long-term divergence exponent and the temporal structure of step-to-step perturbations: a simulation study in two passive walking models}
\author{Philippe Terrier\\[2pt]
  \small Haute Ecole Arc Sant\'e, HES-SO University of Applied Sciences and Arts Western Switzerland\\
  \small Espace de l'Europe 11, 2000 Neuch\^atel, Switzerland\\
  \small \texttt{philippe.terrier@he-arc.ch}}
\date{}

\begin{document}

\maketitle

\begin{abstract}
\noindent\textbf{Purpose.} Divergence exponents are used to assess the local dynamic stability of gait. Human experiments have related the long-term exponent ($\lL$) to the complexity of gait fluctuations. Model studies used perturbations without controlled temporal structure. The question was whether $\lL$ responds to the temporal structure of perturbations in walkers without neuromuscular control.
\textbf{Methods.} Two passive walkers, M1 (pilot) and M2 (exploratory), received step-to-step perturbations of five amplitudes ($\sigma$) and nine temporal structures (ARFIMA parameter $d$), and human stride series of free and metronome walking.
\textbf{Results.} At constant amplitude, $\lL$ increased with $d$ in both models. It was higher with free series than with metronome series or with surrogates. It did not follow $\sigma$ in M1 and followed $\sigma$ in M2. Its double dissociation from the short-term exponent was supported in M1 only. In a descriptive analysis, the condition means of $\lL$ correlated with the scaling exponent of the step durations ($r = 0.980$ in M1 and 0.906 in M2).
\textbf{Conclusion.} In two passive walkers with fixed mechanics, $\lL$ followed the temporal structure of the perturbations; its specificity depended on the model. This response supports the name attractor complexity index (ACI), proposed for $\lL$ from human experiments.

\end{abstract}

\section{Introduction}
\label{sec:introduction}

Divergence exponents measure how fast neighboring trajectories of a dynamical system separate in its state space \citep{rosenstein1993practical}. In gait analysis, two exponents are computed from the same mean divergence curve: the short-term exponent $\lS$, over the first stride or less, and the long-term exponent $\lL$, over several strides \citep{dingwellNonlinearTimeSeries2000,dingwell2006lyapunov}. Simulations of walking models established $\lS$ as a measure of local dynamic stability. When random perturbations of increasing amplitude were applied at each step, $\lS$ increased with the amplitude \citep{su2007dynamic}, and in a three-dimensional model it rose before the fall rate did \citep{roosInfluenceSimulatedNeuromuscular2011}. In a walker whose global stability was varied through its foot radius and hip stiffness, $\lS$ decreased as the largest tolerated perturbation increased, as expected of a measure of gait stability \citep{bruijn2012maximum}. In these studies, $\lL$ did not follow the amplitude of the perturbation \citep{su2007dynamic}, and the strength of its correlation with the largest tolerated perturbation varied across manipulations \citep{bruijn2012maximum}. \citet{bruijnAssessingStabilityHuman2013a} therefore rated the predictive validity of $\lL$ as negative in models and in observational studies.

Experimental studies of human gait relate $\lL$ to the temporal structure of the stride intervals. A systematic review of 62 studies concluded that the evidence supports reading $\lL$ as a measure of the complexity of gait fluctuations \citep{torrent2026stability}. In its meta-analysis, $\lL$ correlated with the scaling exponent of detrended fluctuation analysis (DFA; $r = 0.64$; five studies, 189 participants). Among the experiments that the review synthesizes, those of rhythmic cueing act on the temporal structure of the stride intervals. Stride intervals of human walking fluctuate with long-range correlations \citep{hausdorff1997altered,terrier2005gps}; these fluctuations become anti-persistent when steps are synchronized with a metronome \citep{delignieresFractalDynamicsHuman2009,terrier2012persistenta}. Metronome walking changed $\lL$ more than $\lS$ \citep{terrierNonlinearDynamicsHuman2013a,terrier2019complexity}. Walking with auditory or visual cues also lowered $\lL$ in healthy adults on a treadmill \citep{terrier2019complexity}, and metronome walking lowered it in older and in young adults over ground \citep{piergiovanniEffectsMetronomeWalking2024a,piergiovanniValidityLinearNonlinear2024a}. \citet{terrierMaximumLyapunovExponent2018a} replaced the stride intervals of real trunk accelerations with series of known temporal structure. The type of series explained 52 to 72\% of the variance of $\lL$, depending on the fitting window, and 4\% of the variance of $\lS$. The authors named $\lL$ the attractor complexity index (ACI) and read it as an index of the complexity of gait fluctuations; later work relates the ACI to the automaticity of gait control \citep{terrier2026accelerometrybased,torrent2026stability}.

The walkers of \citet{su2007dynamic} and of \citet{bruijn2012maximum} have stable limit cycles, which noise perturbs at each step. When only the perturbation changes, the mechanics of the walker are the same in all conditions, and a response of the ACI or of $\lS$ is therefore a response to the perturbation. The question is then which property of the perturbation each exponent follows.

The model studies and the experiments on human gait have not been brought together. The model studies of $\lS$ perturbed each step with a random value from a uniform or a normal distribution \citep{su2007dynamic,bruijn2012maximum,williamsPredictingFallRisk2023a}, or added independent noise to a step controller \citep{roosInfluenceSimulatedNeuromuscular2011}. These perturbations differed in amplitude, and none had a controlled temporal structure, the property to which the ACI responded in human data. Other models reproduced the correlations of human stride intervals, but these studies used DFA and computed no divergence exponent \citep{gates2007possible,dingwellHumansOptimallyExploit2010a,ahn2013longrange}. In a limit-cycle model of rhythmic movement, an injected fractal series transferred its structure to the periods of the model \citep{delignieresFractalModelsEventbased2008}. In the simplest walker driven by a chaotic signal, the Lyapunov exponent came close to that of human stride signals \citep{montazerimoghadam2018simple}. To our knowledge, however, no study has injected perturbations of controlled temporal structure into a walking model at constant amplitude and compared the responses of the short-term and long-term divergence exponents.

We asked whether the ACI responds to the temporal structure of step-to-step perturbations in walking models with fixed mechanics. We injected perturbation series into two passive walking models: the simplest walker \citep{garcia1998simplest}, named M1 here, in the protocol of \citet{su2007dynamic}, and the walker with arced feet and a hip spring of \citet{bruijn2012maximum}, named M2. A crossed design varied the amplitude of the series and its temporal structure. The structure was set by the fractional differencing parameter $d$ of autoregressive fractionally integrated moving average noise, ARFIMA(0,\allowbreak$d$,\allowbreak0). We tested two hypotheses in each model. H2, on temporal structure: at constant amplitude, the ACI increases with $d$ and $\lS$ does not follow $d$. H1, on amplitude: under white noise, $\lS$ increases with amplitude and the ACI does not follow amplitude. Together, H1 and H2 define a double dissociation, which tests the specificity of both responses. As a secondary test, we injected stride series recorded in young adults who walked with and without a metronome \citep{hausdorff1996fractal,hausdorff2001longterm}. We made two predictions on these series. S1, on metronome walking: the ACI is higher with each free series than with the metronome series of the same participant and speed. S2, on the order of the values: the ACI is higher with each free series than with surrogates made of its values in random order. Two descriptive analyses were designed after the results. The first related the ACI to the scaling exponent of the step durations of the models. The second described the joint response of $\lS$ and of the ACI to amplitude and temporal structure.

\section{Methods}
\label{sec:methods}

\subsection{Study design}
\label{sec:study_design}

This is a simulation study of two published passive walking models. It involved no participants; the only human data are stride intervals of a public database (Section~\ref{sec:perturbations}). It is the first part of a funded project on the validation of the ACI by simulation, whose analysis plan covers four walking models (Funding). The study had three parts in each model. (1) Replication runs repeated the protocol of the source study, to check the implementation. (2) A crossed design varied the amplitude and the temporal structure of the perturbations, for the tests of H1 and H2. (3) Human stride series were injected for the secondary predictions S1 and S2.

The analysis plan fixed the design, the outcomes, the statistical models and the decision rules. It was registered on OSF Registries on 6 November 2026 \citep{terrier2026simaci}. Two updates followed on 6 and 7 November; the first fixed details of implementation, and the second concerns an analytic predictor that this article does not use. The simulations of the crossed design and of the human series (the campaigns of this study) ran on 7 November 2026, after the registration and its updates. The replication runs preceded the registration, and the plan states their results.

No test of this article is confirmatory. The tests in M1 are pilot tests, because the thresholds of the decision rules were set on a feasibility study of the same model (Section~\ref{sec:statistics}). The tests in M2 are exploratory, because M2 did not meet the replication criteria, which were stated before its replication runs (\appref{app:m2}). The plan reserves confirmatory tests for two further models, which a second article will report. Two descriptive analyses of the ACI were designed after the results had been seen and lie outside the plan (Section~\ref{sec:exploratory}). The plan also covers analyses that this article does not report; supplementary Table S6 lists every registered analysis and the place where it is reported.

\subsection{Walking models and replication runs}
\label{sec:models}

The first model, M1, is the simplest walking model of \citet{garcia1998simplest}: a point mass at the hip, massless legs of unit length, and an inelastic collision at foot strike that exchanges the legs. The model is dimensionless and walks down a slope of 0.009~rad. Its state consists of the stance-leg angle $\theta$, the angle $\phi$ between the legs, and their rates. As in \citet{su2007dynamic}, the perturbation of step $k$ enters the foot-strike condition, $\phi - 2\theta = \varepsilon_k$. The transition map of their Eq.~3 is applied at the perturbed contact. \appref{app:m1} describes the integration and gives the reasons for this reading of the perturbed collision.

The second model, M2, is the walker with arced feet and a hip spring of \citet{hobbelen2007disturbance}, as used by \citet{bruijn2012maximum}. It adds two features to M1: feet shaped as arcs of radius $r_f$ that roll without slip, and a torsional spring of stiffness $k$ at the hip. The equations of motion and the impact map were derived from the laws of classical mechanics. They reduce to those of M1 when $r_f = 0$ and $k = 0$, and they differ from those printed by \citet{hobbelen2007disturbance} in two places (\appref{app:m2}). The perturbation is the height of the floor at each foot strike, $10^{-5} n$, where $n$ is a normal random number \citep{bruijn2012maximum}; we took $n$ as standard normal and independent across steps. The replication runs start on a slope of 0.004~rad (\appref{app:m2}). The crossed design keeps this slope and uses one configuration, a foot radius of 0.45 and a hip stiffness of 5. These values are the midpoints of the ranges of \citet{bruijn2012maximum}, who varied the two parameters one at a time. The configuration was fixed before the calibration run, by a rule that uses no divergence exponent (\appref{app:m2}).

Each model was first run with the perturbation scheme, the number of levels, the number of simulations and the number of strides or steps per simulation of its source study. For M1, these were independent perturbations drawn from a uniform distribution and scaled by an amplitude $\rho$, six amplitudes from 0 to $\rho_{\max}$, and 10 simulations of 300 strides per amplitude \citep{su2007dynamic}. The largest amplitude $\rho_{\max}$ was 0.020~rad, compared with 0.10~rad in the source study (\appref{app:m1}). For M2, they were the four manipulations of global stability of \citet{bruijn2012maximum}, with 10 simulations of 200 steps per configuration. \appref{app:m1} states the replication criteria.

\subsection{Perturbation series}
\label{sec:perturbations}

In the crossed design, the perturbation series were Gaussian ARFIMA(0,\allowbreak$d$,\allowbreak0) series, generated exactly. Independent Gaussian values were multiplied by the Cholesky factor of the covariance matrix of the process, whose autocorrelations follow \citet{hosking1981fractional}. Each realization was centered and rescaled to the standard deviation $\sigma$ over the analyzed steps (\appref{app:testbed}). The parameter $d$ thus changes the autocorrelation of the series and leaves its amplitude unchanged. Each model received one value per step, at the injection point of its source study (Table~\ref{tab:settings}). Nine levels of $d$ covered the anti-persistent and persistent ranges: $-0.30$, $-0.20$, $-0.10$, 0, 0.10, 0.20, 0.30, 0.40 and 0.45. The level $d = 0$ gives Gaussian white noise. Five levels of $\sigma$ were equally spaced from $0.2\smax$ to $\smax$. In each model, $\smax$ was set by a calibration run (\appref{app:testbed}). It was the value of the calibration grid just below the first value at which the fall rate exceeded 5\% at one or more levels of $d$. The middle level, $\sref = 0.6\smax$, was used for all analyses at constant amplitude. Each cell held 10 replicates of 300 strides after the transient (Table~\ref{tab:settings}). The five cells at $d = 0$, which H1 uses, held 20 replicates (\appref{app:settings}). This gives 500 simulations per model (Figure~\ref{fig:design}).

\FigDesign

A simulation that fell before the planned number of strides was redrawn with the seed of the next attempt, up to 20 times. The rules for falls and for numerical failures of the integrator are in \appref{app:testbed}.

The secondary test used the stride series of the database Long-term Recordings of Gait Dynamics, version 1.0.0, on PhysioNet \citep{hausdorff2001longterm,hausdorff1996fractal,pollard2026physionet}. Ten healthy young men walked for 1~h at their usual, slow and fast speeds. They then walked for 0.5~h at each speed with a metronome that beat twice per mean stride interval of the corresponding free walk, each heel strike on a beat \citep{delignieresFractalDynamicsHuman2009}. The database thus holds 30 pairs of free and metronome series. The first 1\,200 strides of each series were kept, and no interval was removed. Each series was centered, divided by its standard deviation and multiplied by $\sref$. All series then had the same amplitude and differed in temporal structure and in the shape of their distribution. Each stride value perturbed one step of the model. The exponents were computed over the 1\,200 steps, or 600 strides of the model. During the transient, the model received the last values of the same 1\,200, in their order; this circular start keeps the temporal structure. For each series, 10 random permutations with fixed seeds served as surrogates: they keep the values and remove their order. A series that made a model fall was excluded with its pair (\appref{app:testbed}).

\subsection{Divergence exponents and other outcomes}
\label{sec:estimators}

Divergence curves were computed by the implementation of the Rosenstein algorithm (Rosenstein engine) of the R package RACING, version 1.0.2, used without modification \citep{rosenstein1993practical,terrier2026racing_article,terrier2026racing_misc}. The state space consisted of the two angles of each model and their rates, without delay embedding. In both models, no step or stride was normalized to a fixed number of samples; the state series thus keep the variation of the step durations (Table~\ref{tab:settings}). For every sample, the engine finds the nearest neighbor outside an exclusion window and follows the distance $\delta_j(i)$ between the sample and its neighbor over $i$ samples (Figure~\ref{fig:estimation}). The mean divergence curve is
\begin{equation}
  y(i) = \left\langle \ln \delta_j(i) \right\rangle_j ,
  \label{eq:divergence}
\end{equation}
where the mean runs over all pairs $j$ that remain in the series at horizon $i$ (\appref{app:settings}). The short-term exponent $\lS$ and the ACI ($\lL$) are slopes of $y$ over the short and the long fitting windows of the source study of each model (Table~\ref{tab:settings}). In M1, the slope is the mean of the local slopes of the curve, after smoothing by a Hamming window of 7 samples \citep{su2007dynamic}. In M2, it is a least-squares slope of the unsmoothed curve (\appref{app:m2}). The exponents are rates: the increase of $y$ per unit of time. They are reported in the unit of each source study (Table~\ref{tab:settings}). In M1, time is dimensionless, in units of $\sqrt{l/g}$, where $l$ is the leg length and $g$ the acceleration of gravity \citep{garcia1998simplest}; one step lasts about 3.88 of these units of model time. In M2, time is counted in strides. The unit of the exponents is thus 1/model time in M1 and 1/stride in M2. The reproduction archive also gives the exponents per step.

\FigEstimation

Each simulation also gave the kinematic variability of each of the four state variables, the two angles and their rates \citep{su2007dynamic,terrierKinematicVariabilityFractal2011}. Each stride was resampled to 101 points, so that all strides share one time base. At each point, the standard deviation of the variable was computed across the strides of the simulation; the kinematic variability is the mean of these 101 standard deviations. It measures how much the trajectory differs from one stride to the next, around the mean stride. The state variability index compares a run with the reference cell ($d = 0$, $\sref$). For each state variable, the kinematic variability of the run is divided by the mean kinematic variability of the runs of the reference cell; the index is the mean of these four ratios. An index of 1 is thus the state variability under white noise at the reference amplitude. This index is exploratory and lies outside the registered analysis plan. The step durations are an output of each model, the times between its successive foot strikes; the perturbation series are its input. The scaling exponent $\alpha$ of the step durations, of the stride durations and of each perturbation series came from first-order DFA \citep{pengQuantificationScalingExponents1995}, configured as in \citet{terrier2026persistent} (\appref{app:settings}). The $\alpha$ of the perturbation series (model input) served to check the generator, for which the expected value is $d + 0.5$ \citep{terrier2026persistent}. The $\alpha$ of the step and stride durations (model output) measured how much of the temporal structure appeared in these durations. Supplementary Table S4 gives these exponents and the index of each cell.

\begin{table}[tp]
\centering
\def\baselinestretch{1}\selectfont 
\caption{Settings of the two models. In M1, the angles are referred to the same leg throughout, which keeps them continuous across heel strikes \citep{su2007dynamic}. In M2, the neighbor exclusion and the slope follow the public code of the first author of the source study \citep{bruijn2021localdynamicstability}.}
\label{tab:settings}
\small
\begin{tabular}{@{}>{\raggedright\arraybackslash}p{0.27\linewidth}>{\raggedright\arraybackslash}p{0.32\linewidth}>{\raggedright\arraybackslash}p{0.32\linewidth}@{}}
\toprule
Setting & M1, simplest walker & M2, arced feet and hip spring \\
\midrule
Source study & \citet{garcia1998simplest}; \citet{su2007dynamic} & \citet{bruijn2012maximum} \\
State space & two angles and their rates, referred to the same leg throughout; no embedding & two angles and their rates, referred to the stance leg; no embedding$^\dagger$ \\
Resampling of the state series & none & uniform time grid over the whole run, 100 samples per stride on average \\
Conversion of the windows into samples & by the mean step duration & 100 samples per stride \\
Neighbor exclusion & one stride & half a stride on each side (50 samples)$^\dagger$ \\
Window of $\lS$ & 0 to 0.1 step & 0 to 0.5 stride \\
Window of the ACI & 3 to 4 steps & 4 to 10 strides \\
Slope & mean of local slopes & least squares on the unsmoothed curve$^\dagger$ \\
Injection point & foot-strike condition & height of the foot-strike surface \\
Perturbation values & one per step & one per step \\
Transient & 20 steps & 34 steps \\
Unit of $\sigma$ & rad & leg length \\
Largest amplitude $\smax$ & $3.9 \times 10^{-3}$ & $3.0 \times 10^{-4}$ \\
Unit of the exponents & 1/model time & 1/stride \\
\bottomrule
\multicolumn{3}{@{}l@{}}{\rule{0pt}{2.6ex}$^\dagger$ Setting not reported by the source study.} \\
\end{tabular}
\end{table}

\subsection{Statistical analysis and decision rules}
\label{sec:statistics}

The crossed design has two factors: the amplitude $\sigma$ and the temporal structure $d$ of the perturbation series. The decision on each hypothesis was made separately in each model. The response of an exponent to a factor was measured in two ways. The first was the proportion of variance explained, $\eta^2$, in a one-way linear model with the factor as a categorical variable. The second was the Spearman correlation, over the simulations, between the exponent of a simulation and the level of the factor in that simulation. For the response of $\lS$ to $\sigma$, for example, each of the 100 simulations at $d = 0$ gives one pair: its $\lS$ and its $\sigma$.

The absence of a response was measured by a range ratio, computed from condition means. A condition mean is the mean of an exponent over the replicates of a cell. The numerator of the ratio is the range of the condition means over the levels of the factor that the exponent is expected to ignore. Its denominator is their range over the levels of the other factor. For the ACI, which is expected to ignore $\sigma$, the numerator is the range over the five levels of $\sigma$ at $d = 0$. The denominator is the range over the nine levels of $d$ at $\sref$. A ratio of 0.25 means that the ACI varied four times less over $\sigma$ than over $d$.

H1, the hypothesis on amplitude, used the five cells at $d = 0$. H1a held when $\lS$ increased with $\sigma$: $\eta^2 \geq 0.50$ and a positive Spearman correlation whose 95\% interval excluded zero. H1b held when the ACI did not follow $\sigma$: a range ratio of the ACI of at most 0.25. H2, the hypothesis on temporal structure, used the nine cells at $\sref$. H2a held when the ACI increased with $d$, under the criteria of H1a. H2b held when $\lS$ did not follow $d$: a range ratio of $\lS$ of at most 0.25, with its range over $d$ at $\sref$ as numerator. A hypothesis was supported when its two parts, a and b, met their criteria. Its support was called robust when the bounds of the intervals also met them (\appref{app:settings}); robustness is reported as a measure of strength. The double dissociation held in a model when H1 and H2 were both supported (Table~\ref{tab:decisions}). A secondary analysis applied each criterion in every slice of the crossed design, that is, in the cells at one fixed level of $d$ or of $\sigma$ (\appref{app:settings}; supplementary Table S5). Table~\ref{tab:decisions} gives the number of slices in which each criterion was met.

Sampling variation alone gives a positive range ratio. A simulated design gave the ratio that is expected in M1 when the ACI does not respond to $\sigma$, the case of H1b (Section~\ref{sec:res_h1}). This design used the within-cell standard deviations of a feasibility study of M1 and 20 replicates per cell (\appref{app:settings}). No such value was computed for M2. Confidence intervals were 95\% percentile intervals from 2\,000 bootstrap resamples of the replicates in each cell (\appref{app:settings}). $p$-values, given in the reproduction archive, did not enter the decisions, because the number of simulations is set by the design. No correction for multiplicity was applied.

Two differences of the ACI were computed for the human series. S1 compared the free and the metronome series of each pair; S2 compared each original free series with the mean of its 10 surrogates. A linear mixed model with a random intercept for the participant, fitted by maximum likelihood, estimated each mean difference. The 95\% profile-likelihood interval of each mean difference was primary. A prediction was supported when this interval excluded zero in the predicted direction. \appref{app:settings} describes three sensitivity analyses of S1 and S2.

The thresholds of these rules were proposed after the results of the feasibility study of M1 had been seen. The ratio of 0.25 and the replication tolerance of 0.10 (\appref{app:m1}) were set so that M1 met them, and no source independent of this feasibility study justifies these values. Each result is therefore stated for each model, without a claim beyond the two models. Every bootstrap and every permutation used a fixed seed.

\subsection{Descriptive analyses of the ACI}
\label{sec:exploratory}

Two descriptive analyses of the ACI were designed after the results of the registered tests had been seen. The first analysis related the ACI to the scaling exponent $\alpha$ of the step durations of the model. The Pearson correlation of these two variables was computed at three levels in each model. Between the 45 condition means, the correlation measures the relation across conditions. Between the 500 runs, it combines the differences between cells with the differences between the runs of a cell. Within cells, a single correlation was computed over the 500 runs. Before this computation, the condition means of the ACI and of $\alpha$ in the cell of a run were subtracted from the values of that run. This subtraction removes the differences between cells; the correlation then measures whether, among runs with the same $d$ and $\sigma$, a higher $\alpha$ goes with a higher ACI. The 95\% intervals were percentile bootstrap intervals, computed as in Section~\ref{sec:statistics}.

The second analysis described the joint response of $\lS$ and of the ACI to amplitude and temporal structure. Each of these two exponents was regressed on $\sigma/\smax$ and on $d$ over the 500 runs, in a linear regression without interaction, with coefficients $b_\sigma$ and $b_d$. In this regression, an increase of $\sigma$ from $0.2\smax$ to $\smax$ changes the exponent by $0.8\,b_\sigma$, and a change $\Delta d$ of $d$ changes it by $b_d\,\Delta d$. The two changes cancel when $\Delta d = -0.8\,b_\sigma/b_d$; this value is the change of $d$ that offsets the increase of the amplitude. A small offsetting change means that the exponent responds little to the amplitude, relative to its response to $d$. The intervals of this analysis were computed in the same way.

Figure~\ref{fig:grid} also gives each condition mean as a percent difference from the condition mean at $d = 0$ of the same amplitude. The 95\% interval of each percent difference came from the same bootstrap, with the two cells resampled independently. These percent differences were added after the results, outside the registered plan; they describe the data and enter no decision. No correction for multiplicity was applied to their 160 intervals, which the reproduction archive gives.

\subsection{Testbed and reproducibility}
\label{sec:testbed}

The models, perturbation ports, estimators and run driver form an R package, walkbench (version 0.6.2), distributed on GitHub and archived on Zenodo (\citealp{terrier2026walkbench}; \appref{app:testbed}). The code was written with generative language models (Claude, Anthropic). Automated tests compare its outputs with the archived outputs of the feasibility study, which it reproduces at full double precision (\appref{app:testbed}). The campaigns of the crossed design and of the human series ran in R 4.5.1 with walkbench 0.6.2, on one computer under Windows (x86-64), with six parallel R processes. The calibration and replication runs used earlier versions of the package, which their logs record. The campaign scripts, the results, the run logs and the scripts that regenerate every figure and table of this article are in a reproduction archive on Zenodo, \url{https://doi.org/10.5281/zenodo.24365284} (Data and code availability).

\section{Results}
\label{sec:results}

\subsection{Replication and manipulation checks}
\label{sec:res_checks}

The simplest walker (M1) met the criteria of partial replication. The walker with arced feet (M2) did not meet the replication criteria, because of the relations of the long-term exponent (\appref{app:m1} and \appref{app:m2}). In the crossed design of each model, all 500 simulations had outcomes, without a fall or a redraw, and every cell entered the analyses. From $d = -0.30$ to 0.45, the mean scaling exponent $\alpha$ of the perturbation series rose from 0.249 to 0.925 in M1, and from 0.249 to 0.926 in M2. The mean $\alpha$ of the step durations of the model rose from 0.168 to 0.857 in M1, and from 0.271 to 0.800 in M2. At constant amplitude of the perturbation series ($\sref$), the mean state variability index changed with $d$ (supplementary Table S4). From $d = -0.30$ to 0.45, the index decreased from 1.074 to 0.673 in M1 and increased from 0.787 to 2.224 in M2.

The metronome series of one participant at two speeds made both models fall, so these two pairs were left out of the tests on the human series, S1 and S2 (Table~\ref{tab:decisions}; supplementary Table S7). In M1, 3 surrogates of metronome series fell and were replaced. The mean $\alpha$ was 0.973 for the free series and 0.313 for the metronome series. For the step durations of the model, it was 0.911 and 0.288 in M1, and 0.786 and 0.325 in M2.

The replication runs of M2 followed the four manipulations of the source study, under the same white noise (\appref{app:m2}). The foot radius was increased (manipulation 1), also with the slope adjusted to keep the step length (manipulation 2). The hip stiffness was increased (manipulation 3), also with the adjusted slope (manipulation 4). In every manipulation, $\lS$ decreased as the logarithm of the largest tolerated perturbation increased, as published (reproduced Pearson correlation $r$ from $-$0.807 to $-$0.987; Table~\ref{tab:m2targets}). For the ACI, the reproduced $r$ was 0.902 [0.774, 0.928], $-$0.191 [$-$0.430, 0.164], 0.261 [0.119, 0.357] and $-$0.141 [$-$0.378, 0.166] in manipulations 1 to 4, and 0.405 [0.298, 0.451] in the pooled data. In manipulation 1, the relation is strong and positive, as in the source study (published $r$ of 0.934). In manipulations 2 and 4, the reproduced relation of the ACI does not have the published direction, and its interval includes zero. Under the rules stated before these runs, M2 is therefore classified as not replicated.

\subsection{Response of the ACI to the temporal structure of the perturbations (H2a)}
\label{sec:res_h2}

Hypothesis H2 stated that, at constant amplitude ($\sref$), the ACI increases with $d$ (H2a) and $\lS$ does not follow $d$ (H2b). In both models, the ACI increased with $d$ at $\sref$, and H2a was met by the point estimates and by the bounds of their intervals (Table~\ref{tab:decisions}). The $\eta^2$ of the ACI on $d$ was 0.719 [0.667, 0.806] in M1 and 0.710 [0.647, 0.819] in M2. A condition mean is the mean of an exponent over the replicates of one cell of the crossed design (Figure~\ref{fig:design}d). From $d = -0.30$ to 0.45, the condition mean of the ACI at $\sref$ rose from 0.0041 to 0.0473 in M1 (unit: 1/model time), and from 0.0168 to 0.0699 in M2 (unit: 1/stride). In the secondary analysis, H2a was also met in each of the 5 slices at fixed $\sigma$, in both models (Table~\ref{tab:decisions}).

\begin{table}[tp]
\centering
\def\baselinestretch{1}\selectfont
\caption{Decisions on the hypotheses of the crossed design (H1, H2) and on the predictions for the human stride series (S1, S2), in the simplest walker (M1) and in the walker with arced feet and a hip spring (M2). Estimates with their 95\% intervals (bootstrap within cells for H1 and H2, profile likelihood for S1 and S2). Range ratio: range of the condition means of an exponent over the factor that it should not follow, divided by their range over the other factor. H1 and H2 are supported when their two parts, a and b, meet their criteria; the support is robust when the bounds of the intervals also meet them. Secondary analysis: slices of the crossed design, at one fixed level of $d$ or of $\sigma$.}
\label{tab:decisions}
\footnotesize
\setlength{\tabcolsep}{5pt}
\renewcommand{\arraystretch}{1.3}
\begin{tabular}{@{}>{\raggedright\arraybackslash}p{0.33\linewidth}>{\raggedright\arraybackslash}p{0.21\linewidth}ll@{}}
\toprule
Hypothesis and criterion & Requirement & M1 (pilot) & M2 (exploratory) \\
\midrule
\multicolumn{4}{@{}l}{\textbf{H1, amplitude}: cells at $d = 0$} \\
H1a: $\lS$ increases with $\sigma$ & $\eta^2 \geq 0.50$ & 0.964 [0.956, 0.974] & 0.882 [0.859, 0.913] \\
 & Spearman $\rho > 0$, interval above 0 & 0.967 [0.956, 0.977] & 0.918 [0.892, 0.941] \\
H1b: the ACI does not follow $\sigma$ & range ratio $\leq 0.25$ & 0.142 [0.079, 0.289] & 0.315 [0.206, 0.461] \\
\textit{Decision on H1} & & \textbf{supported} & not supported \\
\addlinespace[6pt]
\multicolumn{4}{@{}l}{\textbf{H2, temporal structure}: cells at $\sref$} \\
H2a: the ACI increases with $d$ & $\eta^2 \geq 0.50$ & 0.719 [0.667, 0.806] & 0.710 [0.647, 0.819] \\
 & Spearman $\rho > 0$, interval above 0 & 0.841 [0.787, 0.885] & 0.831 [0.778, 0.879] \\
H2b: $\lS$ does not follow $d$ & range ratio $\leq 0.25$ & 0.152 [0.110, 0.218] & 2.017 [1.832, 2.227] \\
\textit{Decision on H2} & & \textbf{supported, robustly} & not supported \\
\addlinespace[6pt]
\textbf{Double dissociation}: H1 and H2 & & \textbf{supported} & not supported \\
Cells retained & & 45/45 & 45/45 \\
\addlinespace[6pt]
\multicolumn{4}{@{}l}{\textbf{Secondary analysis}: slices in which the criterion is met} \\
H1a & 9 slices at fixed $d$ & 9/9 & 9/9 \\
H1b & 9 slices at fixed $d$ & 8/9 & 0/9 \\
H2a & 5 slices at fixed $\sigma$ & 5/5 & 5/5 \\
H2b & 5 slices at fixed $\sigma$ & 5/5 & 0/5 \\
\addlinespace[6pt]
\multicolumn{4}{@{}l}{\textbf{Human stride series}} \\
S1: the ACI is higher with free than with metronome series & mean difference, interval above 0 & 0.026 [0.020, 0.032] & 0.060 [0.053, 0.066] \\
\textit{Decision on S1} & & \textbf{supported} & \textbf{supported} \\
S2: the ACI is higher with free series than with their surrogates & mean difference, interval above 0 & 0.028 [0.024, 0.031] & 0.051 [0.047, 0.055] \\
\textit{Decision on S2} & & \textbf{supported} & \textbf{supported} \\
Pairs retained & & 28/30 & 28/30 \\
\bottomrule
\end{tabular}
\end{table}

Figure~\ref{fig:curves} shows the mean divergence curves: their slope over the long window increased with $d$ in both models. Over the 45 cells of the crossed design, the ACI was higher at $d = 0.45$ than at $d = -0.30$ at each level of $\sigma$ (Figure~\ref{fig:grid}b, d). The tiles of Figure~\ref{fig:grid} give each condition mean as a percent difference from white noise of the same amplitude. For the ACI, the interval of this difference excluded zero in 28 of the 40 tiles in M1 and in 26 of 40 in M2.

\FigCurves
\FigGrid

In a descriptive analysis, the ACI was related to the scaling exponent $\alpha$ of the step durations of the model (Section~\ref{sec:exploratory}; supplementary Figure S1 and Table S1). Between the 45 condition means, the Pearson correlation $r$ of the two variables was 0.980 in M1 and 0.906 in M2 (intervals in supplementary Table S1). Between the 500 runs, $r$ was 0.825 [0.802, 0.847] in M1 and 0.788 [0.763, 0.813] in M2. Within cells, at fixed $d$ and $\sigma$, $r$ was 0.174 [0.090, 0.257] in M1 and 0.345 [0.264, 0.422] in M2. Within cells, the standard deviation of $\alpha$ was 0.069 in M1 and 0.065 in M2, compared with 0.243 and 0.187 between the 45 condition means. That of the ACI was 0.008 in M1 (unit: 1/model time) and 0.013 in M2 (unit: 1/stride), compared with 0.013 and 0.022 between the condition means. The relation between cells therefore comes mostly from the differences between conditions.

\subsection{Human stride series (S1, S2)}
\label{sec:res_human}

On average, the ACI was higher with the free series than with the metronome series of the same participant and speed (S1). It was also higher with the free series than with their surrogates (S2). Both predictions were supported in both models (Table~\ref{tab:decisions}; Figure~\ref{fig:human}). Every interval of S1 and S2 stayed above zero in the three sensitivity analyses. These analyses used a cluster bootstrap of participants, a model with the speed as a covariate, and the first 600 steps of each series.

\FigHuman

The ACI of each original metronome series was also compared with that of its surrogates; this comparison is descriptive and had no prediction. In M1, the mean difference (original minus surrogates) was 0.0018 (unit: 1/model time), with a 95\% interval from $-$0.0032 to 0.0072 that includes zero. In M2, it was $-$0.0105 (unit: 1/stride), with an interval from $-$0.0148 to $-$0.0063 that lies below zero. Supplementary Table S3 gives the $\lS$ obtained with the human stride series, a registered exploratory analysis.

\subsection{Response of the ACI to the amplitude of the perturbations (H1b)}
\label{sec:res_h1}

Hypothesis H1 stated that, under white noise ($d = 0$), $\lS$ increases with the amplitude $\sigma$ (H1a) and the ACI does not follow $\sigma$ (H1b). In M1, the ACI did not follow $\sigma$: the point estimate of the range ratio met the criterion of H1b (Table~\ref{tab:decisions}; Figure~\ref{fig:dissociation}). Without a response of the ACI to $\sigma$, the expected ratio was 0.125 (Section~\ref{sec:statistics}); the observed ratio was 0.142 [0.079, 0.289], and the upper bound of its interval exceeded 0.25. Across the five levels of $\sigma$ at $d = 0$, the condition means of the ACI spanned a range of 0.006 (unit: 1/model time). Across the nine levels of $d$ at $\sref$, they spanned a range of 0.043. In the secondary analysis, H1b was met in 8 of the 9 slices at fixed $d$ (Table~\ref{tab:decisions}). The replication runs of M1 repeated the protocol of \citet{su2007dynamic}, with uniform white noise of six amplitudes $\rho$ (Section~\ref{sec:models}). In these runs, the $r^2$ of the ACI on $\rho$ was 0.090 [0.004, 0.269], compared with 0.010 in the source study (Table~\ref{tab:m1targets}).

\FigDissociation

In M2, H1b was not met, because the ACI also followed $\sigma$: the range ratio was 0.315 [0.206, 0.461] (Table~\ref{tab:decisions}; Figure~\ref{fig:grid}d). Across the levels of $\sigma$, the condition means of the ACI spanned a range of 0.017 (unit: 1/stride), compared with 0.054 across the levels of $d$. In the secondary analysis, H1b was met in none of the 9 slices at fixed $d$, the columns of Figure~\ref{fig:design}d (Table~\ref{tab:decisions}).

A descriptive analysis fitted, for each walking model, a linear regression of the ACI on $\sigma/\smax$ and $d$ over the 500 runs (Section~\ref{sec:exploratory}; supplementary Table S2). In the regression of M1, a decrease in $d$ of 0.046 [0.006, 0.086] offset the increase of $\sigma$ from $0.2\smax$ to $\smax$. In that of M2, the offsetting decrease in $d$ was 0.335 [0.289, 0.382].

\subsection{Short-term exponent and double dissociation (H1a, H2b)}
\label{sec:res_dissociation}

In both models, $\lS$ increased with $\sigma$ at $d = 0$, and H1a was met ($\eta^2 = 0.964$ in M1 and 0.882 in M2; Table~\ref{tab:decisions}; Figure~\ref{fig:dissociation}). H1a was also met in each of the 9 slices at fixed $d$, in both models. In M1, $\lS$ did not follow $d$: H2b was met by the point estimate and by the bound of its interval, with a range ratio of 0.152 [0.110, 0.218]. H2b was also met in each of the 5 slices at fixed $\sigma$. Across the nine levels of $d$ at $\sref$, the condition means of $\lS$ spanned a range of 0.023 (unit: 1/model time), compared with 0.150 across the five levels of $\sigma$ at $d = 0$. In M2, H2b was not met: $\lS$ decreased with $d$, and the range ratio was 2.017 [1.832, 2.227] (Figure~\ref{fig:grid}c). Across the levels of $d$, the condition means of $\lS$ spanned a range of 1.160 (unit: 1/stride), compared with 0.575 across the levels of $\sigma$. H2b was met in none of the 5 slices (Table~\ref{tab:decisions}). For $\lS$, the percent differences from white noise stayed below 3\% in M1, where 19 of the 40 intervals excluded zero; in M2, all 40 intervals excluded zero (Figure~\ref{fig:grid}a, c).

In M1, H1 is supported, without robustness, and H2 is robustly supported, so the double dissociation is supported in this model. This support is not robust, because one bound fails its criterion: the upper bound of the interval of the range ratio of H1b exceeds 0.25 (Table~\ref{tab:decisions}). In M2, H1, H2 and the double dissociation are not supported.
\section{Discussion}
\label{sec:discussion}

We asked whether the ACI responds to the temporal structure of step-to-step perturbations in walking models whose mechanics are fixed. The question follows from the reading of $\lL$ as an index of the complexity of gait fluctuations \citep{terrierMaximumLyapunovExponent2018a,torrent2026stability}. In both models, at constant amplitude, the ACI increased with $d$ (H2a). It was also higher with the free human stride series than with the metronome series and than with surrogates (S1, S2). A second question was the specificity of the responses, which the double dissociation tested (H1 and H2). The double dissociation was supported in the simplest walker (M1) and was not supported in the walker with arced feet (M2). All tests are pilot or exploratory.

\subsection{Response of the ACI to the temporal structure of the perturbations}
\label{sec:aci_structure}

At the reference amplitude, the criteria for an increase of the ACI with $d$ (H2a) were met by the point estimates and by the bounds of their intervals in both models (Table~\ref{tab:decisions}). The criteria of H2a were also met at each of the five amplitudes, in both models (Table~\ref{tab:decisions}). In the primary analysis, the divergence curve averages over fewer pairs of trajectories as the horizon grows, because a pair leaves the average when one of its trajectories reaches the end of the series. A registered sensitivity analysis kept a fixed set of pairs, the same at every horizon; no decision on H1 or H2 changed in either model (\appref{app:settings}; supplementary Table S8).

The predictions on the human stride series (S1, S2) were supported in both models (Table~\ref{tab:decisions}). The models received only the stride intervals, rescaled to one amplitude. The surrogates of S2 differ from their originals only in the order of the values, so the difference in ACI of S2 comes from the temporal structure of the series. The free and the metronome series of S1 also differ in the shape of their distribution (Section~\ref{sec:perturbations}).

\citet{su2007dynamic} and \citet{bruijn2012maximum} perturbed each step of their walkers with random values that had no controlled temporal structure. In their simulations, $\lL$ showed no consistent relation with the manipulations. \citet{terrierMaximumLyapunovExponent2018a} imposed a temporal structure on the stride intervals of recorded trunk accelerations and kept the shape of the signal; the ACI followed the imposed structure. Here, the perturbation series acted on the state of a walker through its mechanics, and the ACI again followed the imposed structure. The scaling exponent of the step durations of the model followed $d$ in both walkers (Section~\ref{sec:res_checks}). A limit-cycle model of rhythmic movement transferred the structure of an injected fractal series to its periods in the same way \citep{delignieresFractalModelsEventbased2008}.

In the meta-analysis of \citet{torrent2026stability}, $\lL$ correlated with the DFA scaling exponent across participants ($r = 0.64$). In the models, the ACI correlated with the scaling exponent $\alpha$ of their step durations, an output of the models, across the condition means ($r = 0.980$ in M1 and 0.906 in M2). Within cells, the correlation was 0.174 in M1 and 0.345 in M2 (Section~\ref{sec:res_h2}). Two properties of the runs of a cell can explain these low values. The runs of a cell received the same $d$ and the same $\sigma$, so their $\alpha$ covered a narrow range. The ACI also varied from run to run at fixed condition (Section~\ref{sec:res_h2}). A narrow range of the variables and a low reliability of their measures both lower a correlation coefficient \citep{piergiovanniEffectsMetronomeWalking2024a}. Human data show the same dependence on the range. In a study of 60 older adults, the correlation between the ACI and $\alpha$ was 0.41 in normal walking and 0.46 in metronome walking, and 0.56 when the two conditions were analyzed together \citep{piergiovanniEffectsMetronomeWalking2024a}. In that study, the test-retest reliability of $\alpha$ was low (intraclass correlation of 0.23). A correlation between the ACI and $\alpha$ thus depends on the range of temporal structures that the sample covers, and on the reliability of the two measures.

\citet{torrent2026stability} proposed that slow fluctuations that are spread over many strides delay the saturation of the divergence curve, which raises $\lL$. \citet{terrierMaximumLyapunovExponent2018a} noted that the divergence curve of the walker of \citet{su2007dynamic} became flat after two strides, and attributed this to the absence of the inter-stride fluctuations of human gait. In both models, the slope of the mean divergence curve over the long window increased with $d$ (Figure~\ref{fig:curves}). This increase agrees with both readings; this study did not test the mechanism.

\subsection{Perturbations in human experiments and in the models}
\label{sec:human_perturbation}

In the eight studies that \citet{torrent2026stability} retained, walking was perturbed by mechanical, visual or other sensory means. The short-term exponent increased in seven of these studies and the long-term exponent decreased in seven, so the two exponents also dissociated in human data. The review relates the decrease of the long-term exponent to an adjustment of motor control: a more deliberate control of gait, as under rhythmic cueing \citep{torrent2026stability}. Passive walkers cannot make this adjustment.

As expected, the passive walkers did not show this decrease: at fixed temporal structure, the ACI did not follow the amplitude in M1 and increased with it in M2 (Figure~\ref{fig:dissociation}). In both models, $\lS$ increased with $\sigma$ (H1a), as the short-term exponent did in the human studies.

The human studies and the models meet on the temporal structure of the fluctuations. Under rhythmic cueing, the stride intervals become anti-persistent \citep{delignieresFractalDynamicsHuman2009,terrier2012persistenta}. If the response of motor control to a perturbation also lowers the persistence of the stride-to-stride fluctuations, its effect corresponds to a decrease in $d$ in the models. This hypothesis agrees with the reading of anti-persistence as the mark of a tightly controlled gait parameter \citep{dingwell2010reinterpreting,roerdink2019tightening}. In treadmill walking, a cognitive task that taxed executive function made the fluctuations of step speed less anti-persistent \citep{decker2013executive}. In the models, a decrease in $d$ lowered the ACI at constant amplitude (H2a). In the linear regression of the ACI, a decrease in $d$ of 0.046 in M1 and of 0.335 in M2 offset the increase of the amplitude over its whole range (Section~\ref{sec:res_h1}). A perturbation that becomes stronger and less persistent corresponds to an increase of $\sigma$ with a decrease of $d$. In the regressions, such a change raises $\lS$ in both models. It lowers the ACI when the decrease in $d$ is larger than the offsetting value, and it raises the ACI when the decrease is smaller (Figure~\ref{fig:grid}). The pattern of the human studies, a higher $\lS$ with a lower ACI, thus appears in M1 for a small decrease in $d$; in M2 it requires a large decrease.

This comparison has three limits. First, the models do not show that the response of motor control to a perturbation changes the temporal structure of gait fluctuations; this step is a hypothesis. Second, such a response can also change the mechanics of gait, which the crossed design holds fixed. Third, a change of the temporal structure does not imply a gain of stability. \citet{mitchellQuantifyingEffectSagittal2022a} simulated a biped with variable joint angles and altered the magnitude and the frequency content of this variability. The fall risk depended on the magnitude of the variability; the frequency content of the variability, a temporal property, had little or no effect on falls.

\subsection{The ACI and the stability of the walkers}
\label{sec:aci_stability}

Without perturbation, both walkers have a stable limit cycle (\appref{app:testbed}). In the crossed design, the parameters of each walker were the same in all cells, and the cells differed only in their perturbation series. The unperturbed limit cycle of each walker and the eigenvalues of its step map were therefore the same in all cells. The ACI thus increased with $d$ in walkers whose mechanics did not change. \citet{torrent2026stability} concluded from human studies that $\lL$ does not behave as a measure of stability. At fixed mechanics, the ACI of the models changed with the perturbation alone, which agrees with this conclusion.

In the replication runs of M2, the mechanics changed under the same white noise (Section~\ref{sec:res_checks}). In the four manipulations, $\lS$ decreased as the largest tolerated perturbation increased, the behavior expected of a stability measure \citep{bruijn2012maximum}. The ACI did not show this behavior. In manipulation 1, where the foot radius increased, the ACI increased with the largest tolerated perturbation ($r = 0.902$). \citet{bruijn2012maximum} already noted that this relation was opposite to the one expected of a stability measure: higher values of $\lL$ went with a higher tolerance of perturbations. The relation was weaker when the hip stiffness increased (manipulation 3), and its interval included zero in the two manipulations in which the slope kept the step length constant (manipulations 2 and 4). Under these manipulations, $\lS$ followed the tolerance of perturbations, and the ACI did not follow it consistently. A change of the mechanics alone can thus change the ACI, and its relation with the largest tolerated perturbation depended on the manipulation.

In human walking, three factors may therefore change together: the level of the perturbations, the mechanics that the walker adjusts, and the temporal structure of the stride-to-stride fluctuations that motor control produces. The adjustment of the mechanics is documented. People who expect a slippery floor shorten their strides and land with a flatter foot \citep{cham2002changes,marigold2002strategies}, and a slippery walkway also increases limb stiffness \citep{cappellini2010motor}. Healthy older adults walk with shorter steps, a longer double support and a more flat-footed landing than young adults \citep{winter1990biomechanical}. In older adults, a shorter stride, a lower speed and a longer double support go with the fear of falling \citep{maki1997gait,chamberlin2005does}. The foot radius and the hip stiffness of M2 are mechanical parameters of this kind, and both change the largest tolerated perturbation \citep{bruijn2012maximum}.

The temporal structure and the long-term exponent change in the same situations. Under visual perturbations, the DFA scaling exponent of lateral step placement decreased by 6\% in young and 26\% in older adults, and the long-term exponent by 13\% and 17\% \citep{franz2015advanced,torrent2026stability}. In patients with chronic pain of the lower limbs, whose gait may be cautious, the long-term exponent of free-living walking was lower than in healthy controls, with a small difference in the short-term exponent \citep{terrierMonitoringGaitQuality2017}. Among older adults with a cautious gait, a fractal index of the stride intervals differed between fallers and non-fallers \citep{herman2005gait}. In 60 community-dwelling older adults, the ACI and the DFA scaling exponent were lower in those who reported a recent fall \citep{terrier2026accelerometrybased}.

\subsection{Specificity of the responses: double dissociation and short-term exponent}
\label{sec:specificity}

In the simplest walker (M1), the ACI did not follow the amplitude and $\lS$ did not follow $d$: H1b and H2b were met (Table~\ref{tab:decisions}). In the walker with arced feet (M2), neither criterion was met. The ACI also followed the amplitude, over a range about one third of its range over $d$, and $\lS$ decreased with $d$ over a range about twice its range over $\sigma$ (Figure~\ref{fig:dissociation}). The double dissociation was therefore supported in M1, without robustness, and was not supported in M2. In the secondary analysis, M1 met the criterion of H1b in 8 of the 9 slices and that of H2b in all 5 slices. M2 met neither criterion in any slice. The difference between the two models thus holds over the whole design. The thresholds of the decision rules were set so that M1 met them in its feasibility study (Section~\ref{sec:statistics}); the support in M1 is not independent of that study. A dependence on the model was also reported for other measures. \citet{jinValidationNewPhasedependent2021a} correlated several measures with gait robustness, defined as the largest perturbation that a walker can handle, in two compass walkers. With point feet, as in M1, the local divergence exponent and most phase-dependent measures correlated with gait robustness; with circular feet, the foot shape of M2, they correlated poorly.

In M2, $\lS$ decreased with $d$ at constant amplitude, so H2b was not met. The cause of this response was not identified. Only M2 showed this decrease, and a separate article will examine the responses of $\lS$ in further models.

\subsection{Replication status and behavior of the walker with arced feet (M2)}
\label{sec:m2_status}

The replication runs of M2 served as a check of the implementation; a full replication study of \citet{bruijn2012maximum} was outside the aim of this work. The relations of $\lS$ with the largest tolerated perturbation were reproduced in the four manipulations. The relation of the ACI lacked the published direction in two of them, so M2 was classified as not replicated. The equations had been derived again: the source study prints none, and one equation printed in the article that it cites agrees with the derived one only when a sign is reversed. Two sensitivity analyses left this classification unchanged. The first replaced the derived impact map with the one printed in the cited article. The second estimated the exponents with 48 combinations of settings, and M2 was not replicated with 47 of them (\appref{app:m2}). The origin of the difference from the published relation was not identified. Under the rules written before these runs, the tests in M2 became exploratory, and no setting was chosen for its agreement with the published values. The results in M2 therefore hold for the walker as implemented here, in one configuration; this applies to the results that agree with the hypotheses and to those that do not. M2 also returned to its limit cycle more slowly than M1, and its state variability increased with $d$, while that of M1 decreased (Section~\ref{sec:res_checks}; \appref{app:testbed}). These properties may contribute to the two responses of M2 that departed from the hypotheses: the ACI increased with $\sigma$, and $\lS$ decreased with $d$. This study did not test this explanation. The analysis plan reserves the confirmatory tests of H1 and H2 for two further models.

\subsection{Place of the results in the validation of the ACI}
\label{sec:validation}

The synthesis of \citet{torrent2026stability} assembled the evidence on $\lL$ in three strands. These are its correlation with the DFA scaling exponent, its response to a temporal structure imposed on stride intervals \citep{terrierMaximumLyapunovExponent2018a}, and its dissociation from the short-term exponent. The present results add to the second strand: the ACI responded to an imposed temporal structure in mechanical systems without neuromuscular control, whose equations are known. For the third strand, the results differed between the two models (Section~\ref{sec:specificity}).

With the human stride series, the ACI was lower with the metronome series than with the free series (prediction S1; Section~\ref{sec:res_human}). This difference matches the lower $\lL$ of walking with auditory or visual cues in the human studies of \citet{terrierNonlinearDynamicsHuman2013a}, \citet{terrier2019complexity} and \citet{piergiovanniEffectsMetronomeWalking2024a}, and in the synthesis of \citet{torrent2026stability}. In human walking, a metronome also changes other features of the gait pattern. In the models, only the perturbation series changed, and this change was sufficient to lower the ACI.

M1 and M2 are passive walkers, without neuromuscular control or attention. Later work relates the ACI to the automaticity of gait control \citep{terrier2026accelerometrybased,torrent2026stability}; this relation cannot be tested in passive walkers. Walking models with step-to-step regulation address this link from the side of control: in a powered compass walker, the regulation of speed from one step to the next enlarged the basin of attraction \citep{patil2020tasklevel,patil2022viability}.

\subsection{Limitations}
\label{sec:limitations}

The temporal structure was varied through the parameter $d$ of ARFIMA noise only. Perturbations with short-range correlations may produce a similar response of the ACI; the crossed design did not include them. DFA meets a related difficulty in short physiological series, where its scaling exponent mixes short-range and long-range correlations \citep{terrier2026persistent}. A later article will compare the two kinds of correlations. The design separates amplitude and temporal structure in the perturbation series only. In the state of the walker, the two remain linked: at constant $\sigma$, the state variability index decreased with $d$ in M1 and increased with $d$ in M2 (Section~\ref{sec:res_checks}). A response of an exponent to $d$ can therefore include a response to the amplitude of the fluctuations of the state. The responses of both exponents to $d$ were thus not tested at constant state variability in either model. The perturbation series entered the models from outside. Models with control produce the correlations of human stride intervals by themselves \citep{gates2007possible,dingwellHumansOptimallyExploit2010a,ahn2013longrange}.

Only two models were tested: M1, the pilot model (Section~\ref{sec:study_design}), and M2, which did not meet the replication criteria (Section~\ref{sec:m2_status}). M2 was tested in one configuration, which combines a foot radius and a hip stiffness that the source study varied separately (Section~\ref{sec:models}). The responses of the two exponents in M2 may differ in other configurations. The replication runs cover other configurations, under white noise of a single amplitude (Section~\ref{sec:res_checks}). The range ratio depends on the levels chosen for the two factors. A design with a wider range of $d$, or with a narrower range of $\sigma$, can give a lower ratio for the same response of the ACI. The threshold of 0.25 therefore applies to this design only. The descriptive analyses of the ACI were designed after the results had been seen, outside the registered plan, and had no prediction. The fitting windows differ between M1 and M2, as in their source studies, so the ACI is compared within each model only. The tests on the human series (S1 and S2) rest on a small and homogeneous sample, the 10 healthy young men of one PhysioNet database (Section~\ref{sec:perturbations}).

\section{Conclusion}
\label{sec:conclusion}

Two passive walking models with fixed mechanics were tested: the simplest walker (M1) and a walker with arced feet and a hip spring (M2). In both, the ACI followed the temporal structure of the step-to-step perturbations at constant amplitude. It was lower with the metronome series and with the surrogates of the free series than with the free series. The ACI did not follow the amplitude in M1 and followed it in M2, and its double dissociation from $\lS$ was supported in M1 only. These results are stated for the two models only, and all tests are pilot or exploratory. In a mechanical system, the ACI thus responds to the temporal structure of the perturbations that the system receives, the property that its reading as a complexity index requires. Passive walkers cannot show a change of the temporal structure that comes from the control of gait. A second article will report the same tests in further walking models and the Lyapunov exponents of the walkers computed from their equations.

\section*{Data and code availability}

The testbed is the R package walkbench, released under the MIT license at \url{https://github.com/Ph-Terrier/walkbench}; version 0.6.2 is archived on Zenodo (\url{https://doi.org/10.5281/zenodo.24354489}). The reproduction archive of this article, on Zenodo (\url{https://doi.org/10.5281/zenodo.24365284}), contains the campaign scripts, the results in CSV files and the run logs with their seeds. It also holds the scripts that regenerate every figure and table, and the version of the testbed that produced them. The divergence curves were computed with RACING version 1.0.2 \citep{terrier2026racing_misc}. The human stride series are available from PhysioNet \citep{pollard2026physionet}, record Long-term Recordings of Gait Dynamics, version 1.0.0 \citep{hausdorff2001longterm}, under the Open Data Commons Attribution License v1.0. They are not redistributed: the archive contains a script that downloads them and checks their SHA-256 sums. The analysis plan, version 1.0, and its two updates are registered on OSF Registries (\url{https://doi.org/10.17605/OSF.IO/M74TK}, 6 and 7 November 2026); the registration was made public on 18 November 2026. The feasibility study that preceded this work was deposited on Zenodo on 14 September 2026 (\url{https://doi.org/10.5281/zenodo.22746492}; \citealp{terrier2026preliminary}). The reproduction archive also holds the scripts and outputs of the descriptive analyses of Section~\ref{sec:exploratory}.

\section*{Declaration of competing interest}

The author declares no known competing financial interests or personal relationships that could have appeared to influence the work reported in this paper.

\section*{Funding}

This work was supported by the HES-SO University of Applied Sciences and Arts Western Switzerland, Fonds de recherche et d'impulsions [grant number AGP 145369]. The funder had no role in the study design, in the analysis and interpretation of the data, in the writing of the article or in the decision to submit it for publication.

\section*{Ethics}

This simulation study used a public, anonymized data set; no ethical approval was required.

\section*{Declaration of generative AI and AI-assisted technologies in the manuscript preparation process}

During the preparation of this work, the author used Claude Opus~5.5 and Claude Fable~5.2 (Anthropic) to search and screen the literature, to draft and revise the text of the manuscript and to write the scripts of its figures and tables. The use of language models in the research itself, to write the code, to draft the analysis plan and to review it, is described in Section~\ref{sec:testbed} and \appref{app:testbed}. After using these tools, the author reviewed and edited the content as needed and takes full responsibility for the content of the published article.

\bibliographystyle{plainnat}
\bibliography{p1_references}

\appendix
\counterwithin{table}{section}
\section{Replication of the simplest walker (M1)}
\label{app:m1}

\paragraph{Perturbation} As in \citet{su2007dynamic}, the perturbation of step $k$, $\varepsilon_k$, enters the foot-strike condition, $\phi - 2\theta = \varepsilon_k$ (Section~\ref{sec:models}). In the replication runs, $\varepsilon_k$ was drawn independently at each step from a uniform distribution between $-\rho/2$ and $+\rho/2$. The amplitude $\rho$, in rad, is the width of this interval.

\paragraph{Integration} The equations of motion were integrated by the solver lsodar of the R package deSolve, with event detection and relative and absolute tolerances of $10^{-10}$ and $10^{-12}$ \citep{soetaert2010solving}. The trajectory was sampled every 0.03 time units. Heel strike was detected when $\phi - 2\theta - \varepsilon_k$ reached zero while increasing. This quantity also reaches zero at mid-swing, when the swing foot passes the stance foot (foot scuffing); it is then decreasing, and this crossing was ignored. A run fell when the stance-leg angle reached 1.2~rad in absolute value, or when no heel strike occurred within 15 time units.

\paragraph{Perturbed collision} \citet{su2007dynamic} perturb the foot-strike condition and keep the transition map of their Eq.~3, whose coefficients hold for the unperturbed contact geometry. They do not state how $\varepsilon_k$ enters the collision. Five readings of the perturbed collision were implemented. In the literal reading of Eq.~3, $\varepsilon_k$ changes only the time of the foot strike. Two readings compute the collision at the actual contact geometry, without or with a rotation to the local slope. The fourth reading changes the slope of step $k$ by $-\varepsilon_k/2$, and the fifth sets bump heights of the floor at the contacts. Each reading was simulated 5 times over 620 steps at $\rho = 0.004$, 0.012, 0.020, 0.050 and 0.100~rad. With the literal reading, no simulation fell up to 0.020~rad, and all fell at 0.050 and 0.100~rad. With each other reading, all simulations fell at 0.012~rad and above. The literal reading was therefore kept.

\paragraph{Amplitudes of the replication runs} \citet{su2007dynamic} simulated amplitudes $\rho$ up to 0.10~rad, the largest that their walker tolerated. In our implementation, the model fell at 0.050~rad and above. The largest amplitude of the replication runs, $\rho_{\max}$, was therefore set at 0.020~rad, the largest that the model tolerated without falls. The source study converts the perturbation into two physical sizes: the height of a bump or dip of the floor under the foot, and an equivalent change of slope. At $\rho = 0.10$~rad, it reports a bump or dip of approximately 0.005 leg lengths and a maximum change of slope of about 0.019~rad. These values were compared with those that the formulas of the same article give on the limit cycle of M1, to check that both implementations use the same scale of perturbation. At the largest perturbation, $\rho/2$, the formulas give 0.0099 leg lengths and 0.025~rad. At the mean absolute perturbation, $\rho/4$, they give 0.0050 leg lengths and 0.0125~rad. The article does not state which perturbation its values describe, so the comparison did not confirm the scale. The replication was therefore judged on the relations with $\rho$: which outcome increases with $\rho$, and which stays unchanged.

\paragraph{Replication criteria} A model was considered replicated when three conditions held. First, every relation that the source study reports had the published direction. Second, each published $r^2$ or $r$ lay within 0.10 of the reproduced value, or inside its 95\% confidence interval. Third, the fixed point, the step period and the eigenvalues of the step map differed by less than 1\% from the published values, where the source study reports them. For M1, the published values are the expansions of \citet{garcia1998simplest} for the stance angle, its rate and the step period (Table~\ref{tab:m1targets}). Their eigenvalues appear only in a figure and were not compared. For M2, neither source study tabulates these quantities, and the third condition did not apply. A relation reported as absent was reproduced when the reproduced $r^2$, or $\eta^2$ for a factor, was at most 0.10. A model was partially replicated when the first condition held for every relation involving $\lS$ or the ACI and another condition failed. It was not replicated when the first condition failed for a relation involving one of the two exponents; its tests of H1 and H2 were then reported as exploratory. Both kinds of model entered the crossed design with their documented implementation. The tolerance of 0.10, in both rules, was set so that the $\lL$ of M1 met it.

\paragraph{Targets} The fixed point and period of \citet{garcia1998simplest}; a linear increase of $\lS$ with $\rho$ (published $r^2 = 0.881$); no relation of $\lL$ or of the largest Floquet multiplier with $\rho$ (published $r^2 = 0.01$ and $0.024$) \citep{su2007dynamic}.

\paragraph{Reproduced values} Under these rules, M1 is partially replicated. Over $\rho > 0$, $\lS$ gave $r^2 = 0.896$ and $\lL$ gave $r^2 = 0.090$. The largest Floquet multiplier gave $r^2 = 0.540$, where the source study reports no relation. The stance angle, its rate and the period at the fixed point differed from the expansions of \citet{garcia1998simplest} by 0.39\%, 0.43\% and 0.71\%. Table~\ref{tab:m1targets} gives the targets and the reproduced values; the expansions of \citet{garcia1998simplest} were evaluated with the coefficients of their appendix table A.2. The largest Floquet multiplier at 50\% of the stride stayed at 0.40 to 0.41 from 0.0005 to 0.004~rad and rose to 0.55 at 0.020~rad, in the nonlinear range of the model. In simulations added from 0.0005 to 0.002~rad, the multiplier was 0.398 to 0.411 with the mean unperturbed stride as reference stride, and 0.335 to 0.341 with the mean stride of each simulation.

\begin{table}[tp]
\centering
\caption{Replication of M1: published and reproduced values. Top: $r^2$ of each relation with the amplitude $\rho$ in \citet{su2007dynamic} and in the replication runs, with 95\% bootstrap intervals; for the kinematic variability, the source study reports $r^2$ of at least 0.996; a relation reported as absent is reproduced when $r^2 \leq 0.10$. Bottom: fixed point and period against the expansions of \citet{garcia1998simplest}.}
\label{tab:m1targets}
\small
\begin{tabular}{@{}>{\raggedright\arraybackslash}p{0.34\linewidth} l l l l@{}}
\toprule
Relation with $\rho$ & Fit & Published $r^2$ & Reproduced $r^2$ & Criterion met \\
\midrule
kinematic variability, $\theta$ & exponential & 0.996 & 0.951 [0.944, 0.958] & yes \\
kinematic variability, $\dot\theta$ & exponential & 0.996 & 0.950 [0.943, 0.958] & yes \\
kinematic variability, $\phi$ & exponential & 0.996 & 0.951 [0.943, 0.960] & yes \\
kinematic variability, $\dot\phi$ & exponential & 0.996 & 0.949 [0.941, 0.957] & yes \\
$\lS$ & linear & 0.881 & 0.896 [0.867, 0.927] & yes \\
$\lL$ & linear & 0.010 & 0.090 [0.004, 0.269] & yes \\
largest Floquet multiplier & linear & 0.024 & 0.540 [0.377, 0.700] & no \\
\bottomrule
\end{tabular}
\vspace{3mm}
\begin{tabular}{@{}>{\raggedright\arraybackslash}p{0.34\linewidth} l l l l@{}}
\toprule
Quantity & Expansion & Model & Difference & Within 1\% \\
\midrule
stance angle at the fixed point & 0.1995 & 0.2003 & 0.39\% & yes \\
its rate at the fixed point & $-$0.1990 & $-$0.1998 & 0.43\% & yes \\
step period & 3.8552 & 3.8825 & 0.71\% & yes \\
\bottomrule
\end{tabular}
\end{table}

\section{Replication of the walker with arced feet and hip spring (M2)}
\label{app:m2}

\paragraph{Model} Two rigid legs of unit length meet at a point mass at the hip. Each foot is an arc of radius $r_f$ whose center lies on the leg, and the stance foot rolls without slip. The foot masses, placed at the ends of the legs, tend to zero. A torsional spring of stiffness $k$, normalized by the foot mass, acts at the hip. The slope enters through tilted gravity. Foot strike occurs when $(1 - r_f)(\cos\theta - \cos(\theta - \phi)) = h$, with the swing foot descending in front of the stance leg, where $h$ is the height of the floor at the new contact relative to the stance foot \citep{hobbelen2007disturbance}. The heights were drawn independently at each step; \citet{bruijn2012maximum} state neither the variance of their random number nor the independence of successive heights. As in M1, a run fell when the stance-leg angle reached 1.2~rad in absolute value, or when no foot strike occurred within 15 time units.

\paragraph{Departures from the published equations} During a step, the walker moves as a double pendulum: the stance leg rolls over its foot as an inverted pendulum, and the swing leg swings about the hip. The equations of motion were derived with a computer algebra system, from Lagrange's equation for the stance leg and from Newton's second law for the swing foot. The mass of each foot tends to zero, as in the simplest walking model \citep{garcia1998simplest}. The impact map follows from the conservation of angular momentum at foot strike: that of the whole walker about the new contact point, and that of the trailing leg about the hip \citep{mcgeer1990passive,garcia1998simplest}. The derived equations were compared with those printed by \citet{hobbelen2007disturbance}, which \citet{bruijn2012maximum} cite. The equation of the stance leg agrees. The equation of the swing leg agrees only when the sign of its third term is reversed; with the printed sign, it does not reduce to the equation of M1 when $r_f = 0$ and $k = 0$. After a step of the floor, the printed impact map evaluates its denominator at the angle of the stance leg before impact. The conservation of angular momentum about the new contact requires the angle after impact. On level ground, the printed and the derived maps are identical, and the source article does not describe the impact after a step of the floor. Our implementation uses the derived equations and impact map. A sensitivity run repeated manipulations 1 and 2 with the printed map; its results follow the reproduced values below.

\paragraph{Slope of the starting configuration} \citet{bruijn2012maximum} print a slope of 0.04~rad. At this slope the starting configuration, which is the simplest walker, has an unstable cycle; its stable range ends near 0.015~rad \citep{garcia1998simplest}. \citet{hobbelen2007disturbance} studied the foot radius at 0.004~rad, and found no stable cycle above the radius at which \citet{bruijn2012maximum} stopped their first manipulation. The replication therefore used 0.004~rad.

\paragraph{Design of the replication} Four manipulations started from the configuration $r_f = 0$, $k = 0$. The first increased the foot radius by steps of 0.05, and the second did the same with the slope adjusted to keep the step length. The third increased the hip stiffness by steps of 0.5 up to 10, and the fourth did the same with the slope adjusted. Each manipulation stopped at the first level without a stable cycle. Each configuration gave its fixed point, the largest tolerated perturbation, and 10 simulations of 200 steps that started on the fixed point. The largest tolerated perturbation is the sum of the largest step up and the largest step down of the floor at the first step. After such a step, the model must still walk 30 steps on level ground \citep{bruijn2012maximum}. Neither source defines the step length that manipulations 2 and 4 keep constant; the hip displacement along the slope over one step was used.

\paragraph{Estimation settings} \citet{bruijn2012maximum} state the time normalization and the fitting windows of their exponents. They report neither the neighbor exclusion nor the slope method. The primary setting follows the file lds\_calc.m of the public code of the first author \citep{bruijn2021localdynamicstability}; no document states that the source study used this code. The file was read from the public repository, where its last change dates from 10 February 2020; the archived version 1.1 contains the same file. Its header dates the code from 2017. The setting excludes the neighbors within half a stride on each side (50 samples), averages the logarithms of the distances over a shrinking set of pairs, and fits least-squares slopes. The solver returns the trajectory at a fixed time interval, with the exact state at each foot strike. As in the source study, the trajectory was resampled by linear interpolation on a uniform time grid, with 100 samples per stride on average over the run. The angles were referred to the stance leg, so they jump at each foot strike, when the legs exchange their roles. Each grid point was therefore interpolated from the samples of its own step only, which preserves the jumps. The exponents of the replication runs were also estimated with 48 settings, which crossed five choices. The angles were referred to the stance leg or, as in M1, to the same leg throughout. The neighbor exclusion was one stride, one step or the mean period of the series \citep{rosenstein1993practical}. The curve was the mean of the logarithms of the distances or the logarithm of their mean. The set of pairs was shrinking or fixed, and the slope was a least-squares fit or the mean of the local slopes. The setting closest to the source study was designated after the first estimation had been seen and before any statistic of the 48 settings was computed. An earlier designation was based on other papers of the same group. It was replaced before these statistics and remains one of the 48 settings.

\paragraph{Configuration in the crossed design} The replication runs vary one element of M2 at a time. At their starting configuration, with zero foot radius and hip stiffness, the equations of M2 are those of M1. The crossed design therefore uses a foot radius of 0.45 and a hip stiffness of 5, the midpoints of the ranges of \citet{bruijn2012maximum}, on the slope of 0.004~rad. Two conditions must hold: a stable limit cycle, and a largest tolerated perturbation at least equal to that of the starting configuration. If one fails, the foot radius is lowered by steps of 0.05 at a hip stiffness of 5. If no radius meets both conditions, the hip stiffness is lowered by steps of 1 at a foot radius of 0.45. The search computes fixed points, eigenvalues and tolerated perturbations only, and no divergence exponent. This rule was proposed after the exponents of the replication runs had been seen; these exponents do not enter it. Both conditions held at the first values tried, a foot radius of 0.45 and a hip stiffness of 5, and the search stopped there. With these values, the largest eigenvalue modulus of the step map was 0.923, and the largest tolerated perturbation was 0.00816 leg lengths, against 0.00183 at the starting configuration. The configuration was fixed before the calibration run of M2.

\paragraph{Targets} In each of the four manipulations described above (Design of the replication), the source study reports two relations with the logarithm of the largest tolerated perturbation. $\lS$ decreases as it increases (Pearson correlation $r$ from $-0.868$ to $-0.997$), and $\lL$ increases with it ($r$ from 0.040 to 0.934). These values come from Tables 1 and 3 of \citet{bruijn2012maximum}, where the correlation is written $R$.

\paragraph{Reproduced values and status} Under the rules of \appref{app:m1}, M2 is not replicated. In manipulations 2 and 4, the relation of $\lL$ with the largest tolerated perturbation does not have the published direction. The tests of H1 and H2 in M2 are therefore exploratory. The status was the same with the printed impact map and with 47 of the 48 estimation settings; one setting gave a partial replication. Table~\ref{tab:m2targets} gives the published and the reproduced correlations for the primary setting. With each of the 48 estimation settings, the published direction held for the correlation of $\lS$ with the logarithm of the largest tolerated perturbation, in every manipulation and in the pooled data. For the correlation of $\lL$, the published direction held with all 48 settings in manipulations 1 and 3 and in the pooled data. It held with 32 settings in manipulation 2 and with one setting in manipulation 4. This setting combined the angles referred to the stance leg, a neighbor exclusion of one step, the logarithm of the mean distance, a fixed set of pairs and the mean of the local slopes. With it, every relation had the published direction; in manipulation 4, the correlation of $\lL$ was 0.028, against a published value of 0.384. Six relations failed the second condition, which is a partial replication under the rules of \appref{app:m1}. The crossed design used the primary setting, with which M2 is not replicated. In manipulation 1, the largest tolerated perturbation at $r_f = 0.85$ was 0.0319 leg lengths, against about 0.046 read from Fig.~2 of \citet{bruijn2012maximum}. In manipulation 2, the largest tolerated perturbation reached its maximum, 0.0030 leg lengths, at $r_f = 0.70$; the same figure shows a maximum of about 0.008 near $r_f = 0.85$. Our implementation had no stable cycle at $r_f = 0.95$ in manipulation 2, where the same figure shows one. With the printed impact map, the value of manipulation 1 decreased from 0.0319 to 0.0261 leg lengths, and the values of manipulation 2 changed by at most 2\%. The printed impact map therefore does not explain these differences, and their cause was not identified. Our implementation of M2 agreed with each qualitative description that \citet{hobbelen2007disturbance} give of the effects of the foot radius and of the hip stiffness.

\begin{table}[tp]
\centering
\caption{Replication of M2 with the primary estimation setting: Pearson correlation of each exponent with the logarithm of the largest tolerated perturbation across the levels of each manipulation and in the pooled data, published by \citet{bruijn2012maximum} and reproduced, with 95\% bootstrap intervals. Manip.: manipulation of the source study. Same sign: the reproduced correlation has the sign of the published one. Condition 2: the published value lies within 0.10 of the reproduced value or inside its interval.}
\label{tab:m2targets}
\small
\begin{tabular}{@{}l l r l l l l@{}}
\toprule
Exponent & Manip. & Levels & Published $r$ & Reproduced $r$ & Same sign & Condition 2 \\
\midrule
$\lS$ & 1 & 18 & $-$0.997 & $-$0.987 [$-$0.990, $-$0.979] & yes & yes \\
$\lS$ & 2 & 19 & $-$0.967 & $-$0.807 [$-$0.834, $-$0.771] & yes & no \\
$\lS$ & 3 & 21 & $-$0.868 & $-$0.887 [$-$0.899, $-$0.870] & yes & yes \\
$\lS$ & 4 & 21 & $-$0.981 & $-$0.969 [$-$0.973, $-$0.963] & yes & yes \\
$\lS$ & pooled & 79 & $-$0.786 & $-$0.719 [$-$0.729, $-$0.705] & yes & yes \\
$\lL$ & 1 & 18 & 0.934 & 0.902 [0.774, 0.928] & yes & yes \\
$\lL$ & 2 & 19 & 0.675 & $-$0.191 [$-$0.430, 0.164] & no & no \\
$\lL$ & 3 & 21 & 0.040 & 0.261 [0.119, 0.357] & yes & no \\
$\lL$ & 4 & 21 & 0.384 & $-$0.141 [$-$0.378, 0.166] & no & no \\
$\lL$ & pooled & 79 & 0.594 & 0.405 [0.298, 0.451] & yes & no \\
\bottomrule
\end{tabular}
\end{table}

\section{Estimator settings, other outcomes and statistical details}
\label{app:settings}

\paragraph{Divergence curve} The Rosenstein engine of RACING v1.0.2 was used without modification. The initial separation, at $i = 0$, was added to the curve that the engine returns, under the same rules, and the slopes were computed on that curve (Section~\ref{sec:estimators}).

\paragraph{Other outcomes} The kinematic variability of a state variable is the mean, over the 101 points of the normalized stride, of its standard deviation across strides (Section~\ref{sec:estimators}). The largest Floquet multiplier at a Poincaré section of the normalized stride is the largest eigenvalue modulus of the stride-to-stride map of the deviations from the mean stride, fitted by least squares. In M2, the largest Floquet multiplier estimated from the simulated strides was 1 or more in 9 of the 500 simulations of the crossed design (largest 1.054), all at $d$ from 0.20 to 0.45. In M1, the largest estimated multiplier was 0.793.

\paragraph{Set of pairs} The set of pairs shrinks as the horizon grows: a pair leaves the curve when one of its trajectories reaches the end of the series. The sensitivity analysis with a fixed set of pairs was computed for the crossed design (supplementary Table S8). At every horizon, the fixed set holds only the pairs that remain in the series at the end of the long window. With the shrinking set, a median of 98.7\% of the pairs of horizon 0 remained at the end of the long window in M1, and 93.8\% in M2. For this analysis, the simulations were generated again from their seeds on another computer, under R 4.6.0 and Linux; the campaigns ran under R 4.5.1 and Windows. Their $\lS$ differed from the archived values by at most $2.1 \times 10^{-5}$ in M1 and $1.6 \times 10^{-3}$ in M2, in the unit of each source study. The primary analysis of the regenerated simulations gave the archived decisions. On these simulations, the two sets of pairs gave the same decisions.

\paragraph{Other sensitivity analyses} Three sensitivity analyses of S1 and S2 are reported in this article: the cluster bootstrap of participants, the model with the speed as covariate, and the first 600 steps of each human series. For the replication of M2, two sensitivity analyses are reported in \appref{app:m2} (Reproduced values and status): the 48 estimation settings and the printed impact map. No cell was flagged for falls, so the analysis without flagged cells equals the primary analysis. Other sensitivity analyses were registered as exploratory and were not computed for the crossed design (supplementary Table S6). These are the other slope method of each model, uniform white noise at $d = 0$, and other neighbor exclusions and numbers of neighbors. They also include delay embedding, the map of the fitting window and the other injection point.

\paragraph{Detrended fluctuation analysis} Up to 30 box sizes were spaced geometrically from 16 values to one third of the series \citep{terrier2026persistent}. The exponent $\alpha$ was not computed on series shorter than 96 values.

\paragraph{Statistical details} The Spearman correlations were computed over all replicates, with ties at each level of the factor. A hypothesis was robustly supported when the lower bound of the interval of $\eta^2$ was at least 0.50 and the upper bound of the interval of the range ratio at most 0.25. In S1 and S2, a cluster bootstrap of participants (2\,000 resamples, participants relabeled in each, seed 20261106) and a model with the speed as a covariate were sensitivity analyses. That model used sum contrasts, so that its intercept is the mean difference over the three speeds. In the secondary analysis, the range ratio of a slice has its own numerator and the denominator of the primary analysis. For the ACI in the slice at one level of $d$, the numerator is the range of the condition means over the five levels of $\sigma$ at that $d$. The denominator remains their range over $d$ at $\sref$. The slices of an exponent therefore share one reference range. The $\eta^2$ of the factor that should not matter was reported as a descriptive measure, and no $\eta^2$ was corrected for its small-sample bias. In each bootstrap resample, the numerator and the denominator of a range ratio were recomputed from the same resample. The cells of H1 held 20 replicates for the following reason. A simulation of the design used the within-cell standard deviations of the feasibility study of M1 and assumed that the ACI does not follow $\sigma$. With 10 replicates per cell, the range ratio of H1b then exceeded 0.25 in 14\% of the simulated designs, and in 0.9\% with 20 replicates. Every cell, model and criterion is reported, with the failed criteria, the redraws and the hypotheses that could not be tested (supplementary Tables S4, S5 and S7).

\section{The testbed}
\label{app:testbed}

\paragraph{Seeds} The master seed of campaign $k$ of a model was $m = 10^8 + 10^7 k$, recorded in its log; the human series and the crossed design were campaigns 2 and 3. The replication runs of M1 and M2 preceded this rule and kept the master seeds of their scripts: $10^7$ and $1.22 \times 10^7$ in M1, $3 \times 10^7$ in M2. The runs that replay the feasibility study kept its seed rule. The seed of each attempt was $s = m + 100\,000\,a + 100\,c + r$, where $a$ is the number of the attempt, from 0, and $c$ and $r$ are the indices of the cell and of the replicate, from 1. The cells were numbered in the order of the factor levels, with $d$ varying fastest. No seed is therefore shared between campaigns. The calibration runs had master seeds below $10^8$; that of $\sigma$ in M2 used $5 \times 10^7$. The bootstrap used the seed 20261106. Permutation $i$ of human series $j$ used the seed $7 \times 10^7 + 1\,000\,j + i$, with $i$ from 1 to 10 for the surrogates and from 11 upward for their replacements.

\paragraph{Transient} Each simulation started at the fixed point of its model, and the perturbation series was stationary from its first value. The transient let the variance of the state build up to its stationary level. Its length was the smallest number of steps $n$ such that $\lambda^{2n} \leq 0.005$, with a minimum of 20, where $\lambda$ is the largest eigenvalue modulus of the step map. This rule gave 20 steps in M1 ($\lambda = 0.589$) and 34 steps in M2 ($\lambda = 0.923$). In a linear approximation and under white noise, 20 steps in M2 would have left 4\% of the stationary variance missing at the first analyzed step. Each ARFIMA realization included the transient. It was centered and rescaled on the values that followed it, which then had the mean 0 and the standard deviation $\sigma$; the DFA of the perturbation series used the same values. The human series entered after a transient made of their own last values (Section~\ref{sec:perturbations}).

\paragraph{Calibration runs} The calibration run of M2 used its configuration of \appref{app:m2} and the series of the crossed design, with 600 values after the transient. The levels of $\sigma$ rose from $2 \times 10^{-5}$ leg lengths by a ratio of about 1.10. Each level held 10 replicates at each level of $d$, with up to 20 redraws after a fall. At a given $d$, replicate and attempt, all levels of $\sigma$ used the same realization, rescaled. The run stopped three levels after the first level that failed the criterion. With 10 replicates per cell, one fall gives a fall rate of 1/11, so the criterion of 5\% allowed no fall. This criterion passes a level whose true fall rate is 5\% with a probability of 0.60, and a level whose true rate is 10\% with a probability of 0.35. The calibration run of M2 used a transient of 20 steps. A second run of the two levels around $\smax$, with the transient of 34 steps, gave the same fall rate in every cell. The calibration run of M1 followed the same procedure, with a transient of 20 steps and the master seed $4 \times 10^7$. Its first level that failed the criterion was $4.3 \times 10^{-3}$~rad, with 2 falls in 12 attempts at $d = -0.20$; no integration error occurred.

\paragraph{Falls} A simulation that fell at all 21 attempts was recorded with missing outcomes. The fall rate of a cell was its number of falls divided by its number of attempts. A cell with a fall rate above 20\% was flagged and kept in the analyses, and a sensitivity analysis removed the flagged cells. An attempt in which the integrator stopped on an error, without a fall, was a numerical failure. With a generated input, the next attempt drew a new series, and numerical failures counted among the 21 attempts; with a fixed input, a numerical failure was treated as a fall. Numerical failures were not counted as falls, and a cell was also flagged when its rate of falls and numerical failures combined exceeded 20\%. A cell entered the analyses only when at least 80\% of its planned replicates had outcomes, and the statistics used the available replicates of each cell.

\paragraph{Falls with the human series} A human series is a fixed input, so a fall could not be redrawn. A series that made a model fall was excluded with its pair and its surrogates, and the report lists it. A surrogate that fell was replaced by a new permutation with the next seed. When more than five of the 30 pairs were excluded in a model, the experiment was repeated at $0.2\smax$. The repeat then replaced the first run for S1 and S2, and the first run was reported as exploratory. With more than five pairs still excluded, S1 and S2 could not be tested in that model.

\paragraph{Verification} The package was developed from version 0.1.0 to version 0.6.2. Version 0.1.0 passed its automated tests and R~CMD~check. R~CMD~check is the standard check of an R package: it installs the package, runs its examples and its tests, and inspects the code and the documentation for errors and inconsistencies. Run through the package, the 150 simulations of the feasibility study gave the archived values of every outcome and every divergence curve, identical at full double precision. The linearization of M1 at its fixed point was identical to that of a separate implementation. Version 0.2.0 implements the primary settings of M2 and passed the same checks. On 24 replication runs of M2 simulated again, its exponents differed by less than $5 \times 10^{-15}$ from those of the same setting in the grid of \appref{app:m2}. Version 0.3.0 applies the rule of the transient and passed the same checks. The replication runs of M1 and M2 used version 0.1.0, and the sensitivity run with the printed impact map used version 0.1.1. Version 0.6.2 was used for the campaigns of the human series and of the crossed design, in both models. This version passed its 78 automated tests, which contain 1\,709 assertions, and R~CMD~check returned no warning and no note.

\paragraph{Package} In the R package walkbench, each model has a model adapter with its state variables, injection points, fixed point, step map and fall rule. Experiments therefore run on either model without change. The perturbation port applies an input series at a named injection point, one value per step, without rescaling. The run driver computes the seed of each attempt, draws the input, simulates, redraws after a fall, removes the transient and computes the outcomes. The run log of a campaign records its master seed, its design, the version and commit of the package, and the R environment. Code from other repositories is held as pinned copies, unchanged; among them are the walker file of the feasibility study and the Rosenstein engine of RACING v1.0.2. A manifest gives the SHA-256 sum of each copy; the package recomputes the sum before use and stops when the sum differs. Version 0.6.2 contains the two models of this article, M1 and M2. A user guide in the repository of the package describes its functions with worked examples, for readers who wish to run their own simulations. The two further models of the analysis plan will be added to the package.

\paragraph{Use of language models in the research} Generative language models wrote the code and drafted the analysis plan; the author approved every design choice and threshold of the plan (Section~\ref{sec:statistics}). Claude Fable~5 (Anthropic) wrote the first version of the walker code of the feasibility study in August 2026, and Claude Opus~5 changed it in September 2026. Claude Opus~5.5 wrote the testbed, the scripts of the analyses and of the figures, and drafted the analysis plan in November 2026. Claude Fable~5.2 audited the manuscript and wrote further scripts of figures and tables in November 2026. Before the registration, Claude Fable~5.1 (Anthropic) and Gemini~4 Argon (Google) were asked to find ambiguous rules, inconsistencies and gaps in the plan. Before the campaigns, Claude Fable~5.1 also reviewed the code of the campaign scripts. The automated tests described above check the code against the archived outputs of the feasibility study.

\clearpage
\counterwithout{table}{section}
\setcounter{figure}{0}
\setcounter{table}{0}
\renewcommand{\thefigure}{S\arabic{figure}}
\renewcommand{\thetable}{S\arabic{table}}
\section*{Supplementary material}\suppressfloats[t]\renewcommand{\theHfigure}{S\arabic{figure}}\renewcommand{\theHtable}{S\arabic{table}}
\label{sec:supplement}

Tables S1 and S2 report the descriptive analyses of the ACI over the crossed design, and Table S3 gives $\lS$ with the human stride series. Tables S4 and S5 report every cell and criterion of the crossed design, as the registered analysis plan requires. Figure S1 shows the response of the ACI to $d$ at each amplitude and its relation with the scaling exponent of the step durations of the model. Table S6 lists the registered analyses with the place where each is reported. Table S7 counts the simulations with the human stride series, and Table S8 gives the sensitivity analysis with a fixed set of pairs. The reproduction archive holds every value.

\begin{figure}[p]
\centering
\includegraphics[width=\textwidth]{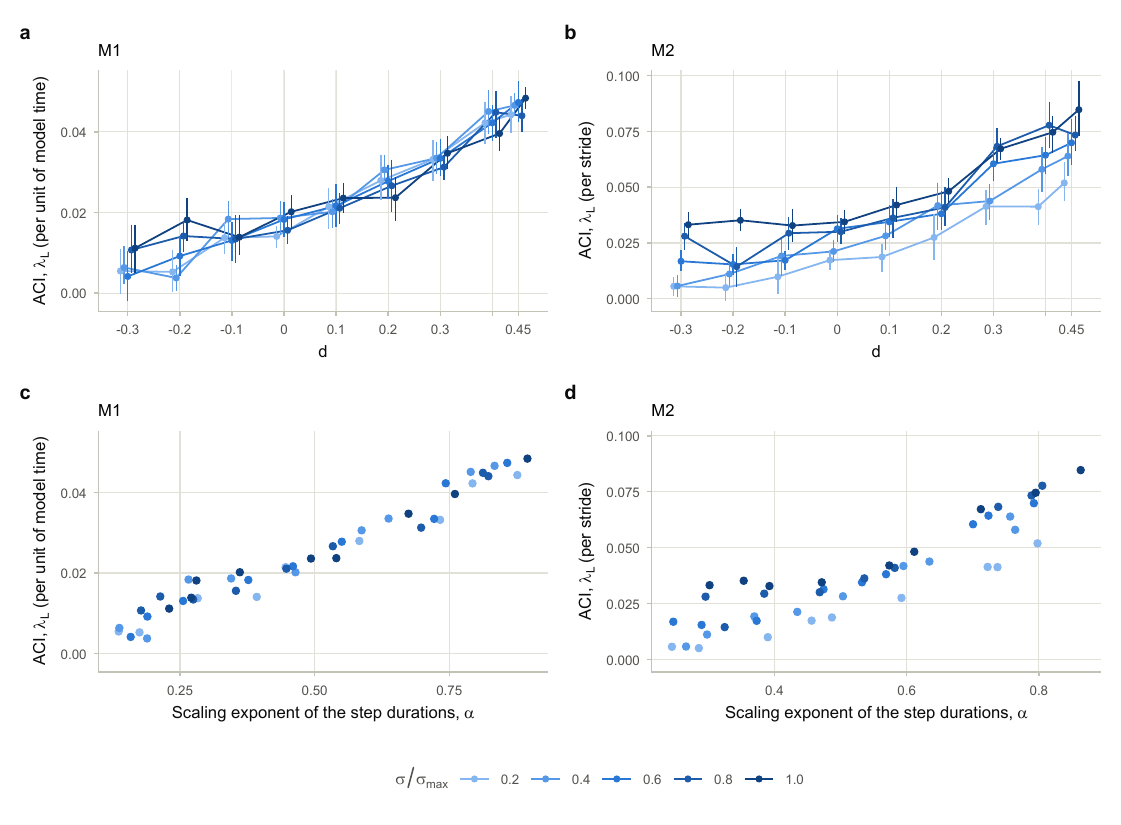}
\caption{Response of the ACI over the crossed design, in the simplest walker, M1 (a, c), and in the walker with arced feet and a hip spring, M2 (b, d). (a, b) Condition means of the ACI against $d$, one line per level of the amplitude $\sigma/\smax$. Bars: 95\% percentile bootstrap intervals (Section~\ref{sec:statistics}). The points of the five levels are offset along $d$ for legibility. (c, d) Condition means of the ACI against the condition means of the scaling exponent $\alpha$ of the step durations of the model, one point per cell (45 per model), colored by the level of $\sigma/\smax$. Within a model, the two panels share their vertical scale. The ACI is per unit of model time in M1 and per stride in M2.}
\label{fig:s2}
\end{figure}

\begin{table}[tp]
\centering
\caption{Relation between the ACI and the scaling exponents $\alpha$ of the detrended fluctuation analysis (DFA) over the crossed design, in the simplest walker (M1) and in the walker with arced feet and a hip spring (M2). Pearson correlation $r$ and Spearman correlation $\rho$ of the ACI with the $\alpha$ of the step durations, of the stride durations and of the perturbation series (the input of the model), with 95\% intervals. Levels: the 45 condition means; the 9 condition means at the reference amplitude $\sref$; the 500 runs; the 500 runs within cells, after subtraction of the condition means of both variables ($\rho$ not computed). Intervals: percentile bootstrap intervals from 2\,000 resamples of the replicates within each cell. For correlations between condition means, the resampling adds sampling variance to the condition means, so that these intervals are shifted toward lower values and can exclude the estimate. Descriptive analysis, outside the registered analysis plan.}
\label{tab:s1}
\small
\begin{tabular}{@{}l l l r l l@{}}
\toprule
Model & Exponent $\alpha$ of & Level & Points & $r$ [95\% interval] & $\rho$ [95\% interval] \\
\midrule
M1 & step durations & condition means & 45 & 0.980 [0.950, 0.976] & 0.977 [0.942, 0.977] \\
 &  & condition means at $\sref$ & 9 & 0.988 [0.950, 0.992] & 1.000 [0.950, 1.000] \\
 &  & runs & 500 & 0.825 [0.802, 0.847] & 0.810 [0.783, 0.834] \\
 &  & runs, within cells & 500 & 0.174 [0.090, 0.257] &  \\
 & stride durations & condition means & 45 & 0.975 [0.941, 0.971] & 0.975 [0.938, 0.974] \\
 &  & runs & 500 & 0.799 [0.773, 0.824] & 0.788 [0.760, 0.814] \\
 & perturbation series & condition means & 45 & 0.966 [0.932, 0.965] & 0.972 [0.929, 0.969] \\
 & (model input) & runs & 500 & 0.801 [0.775, 0.826] & 0.789 [0.761, 0.815] \\
\addlinespace
M2 & step durations & condition means & 45 & 0.906 [0.869, 0.916] & 0.910 [0.860, 0.920] \\
 &  & condition means at $\sref$ & 9 & 0.975 [0.933, 0.985] & 0.983 [0.883, 1.000] \\
 &  & runs & 500 & 0.788 [0.763, 0.813] & 0.780 [0.751, 0.807] \\
 &  & runs, within cells & 500 & 0.345 [0.264, 0.422] &  \\
 & stride durations & condition means & 45 & 0.914 [0.866, 0.921] & 0.927 [0.863, 0.927] \\
 &  & runs & 500 & 0.750 [0.718, 0.782] & 0.749 [0.718, 0.780] \\
 & perturbation series & condition means & 45 & 0.867 [0.827, 0.881] & 0.866 [0.814, 0.885] \\
 & (model input) & runs & 500 & 0.751 [0.724, 0.778] & 0.748 [0.718, 0.779] \\
\bottomrule
\end{tabular}
\end{table}

\begin{table}[tp]
\centering
\caption{Joint response of $\lS$ and of the ACI to the amplitude and to the temporal structure of the perturbations over the crossed design (M1: per unit of model time; M2: per stride). Top: linear regression of each exponent on $\sigma/\smax$ and on $d$ over the 500 runs, without interaction. Coefficients with 95\% intervals, coefficient of determination $R^2$, and gain in $R^2$ of the form with the interaction and the squares of both variables. Bottom: change of $d$ that offsets an increase of $\sigma$ from 0.2 to 1.0~$\smax$, computed from the two coefficients; a negative value is a decrease of $d$. Intervals: 95\% percentile bootstrap intervals from 2\,000 resamples of the replicates within each cell. Descriptive analysis, outside the registered analysis plan.}
\label{tab:s2}
\small
\begin{tabular}{@{}l l l l r r@{}}
\toprule
Model & Exponent & Coefficient of $\sigma/\smax$ & Coefficient of $d$ & $R^2$ & Gain in $R^2$ \\
\midrule
M1 & $\lS$ & 0.178 [0.175, 0.182] & 0.023 [0.019, 0.026] & 0.891 & 0.059 \\
 & ACI & 0.0030 [0.0004, 0.0056] & 0.0518 [0.0487, 0.0547] & 0.671 & 0.029 \\
M2 & $\lS$ & 0.630 [0.609, 0.652] & $-$1.489 [$-$1.519, $-$1.458] & 0.945 & 0.017 \\
 & ACI & 0.0308 [0.0270, 0.0346] & 0.0736 [0.0689, 0.0784] & 0.661 & 0.036 \\
\bottomrule
\end{tabular}
\vspace{4mm}

\begin{tabular}{@{}l l l >{\raggedright\arraybackslash}p{0.38\linewidth}@{}}
\toprule
Model & Exponent & Offsetting change of $d$ & Note \\
\midrule
M1 & $\lS$ & $-$6.306 [$-$7.579, $-$5.412] & outside the range of $d$ of the design \\
 & ACI & $-$0.046 [$-$0.086, $-$0.006] &  \\
M2 & $\lS$ & 0.338 [0.325, 0.353] &  \\
 & ACI & $-$0.335 [$-$0.382, $-$0.289] &  \\
\bottomrule
\end{tabular}
\end{table}

\begin{table}[tp]
\centering
\caption{$\lS$ with the human stride series, over the primary fitting window, in the simplest walker (M1, per unit of model time) and in the walker with arced feet and a hip spring (M2, per stride). Series with outcomes, without the two excluded pairs: the original series of free and of metronome walking, and 10 surrogates of each. Mean, standard deviation (SD), minimum and maximum over the series. First 600: mean of $\lS$ estimated on the first 600 kept steps. Difference: mean of the original series minus mean of their surrogates. Exploratory analysis of the registered plan.}
\label{tab:s3}
\small
\begin{tabular}{@{}l l l r r r r r r r@{}}
\toprule
Model & Condition & Series & $n$ & Mean & SD & Minimum & Maximum & First 600 & Difference \\
\midrule
M1 & free & original & 28 & 0.6502 & 0.0114 & 0.6234 & 0.6689 & 0.5863 & $-$0.0021 \\
 &  & surrogates & 280 & 0.6524 & 0.0094 & 0.6160 & 0.6807 & 0.5873 &  \\
 & metronome & original & 28 & 0.6489 & 0.0113 & 0.6127 & 0.6678 & 0.5853 & 0.0009 \\
 &  & surrogates & 280 & 0.6480 & 0.0120 & 0.5913 & 0.6727 & 0.5837 &  \\
\addlinespace
M2 & free & original & 28 & 2.333 & 0.175 & 2.034 & 2.653 & 2.165 & $-$0.799 \\
 &  & surrogates & 280 & 3.132 & 0.052 & 2.986 & 3.266 & 2.910 &  \\
 & metronome & original & 28 & 3.108 & 0.221 & 2.745 & 3.615 & 2.875 & $-$0.004 \\
 &  & surrogates & 280 & 3.112 & 0.065 & 2.884 & 3.279 & 2.894 &  \\
\bottomrule
\end{tabular}
\end{table}

\clearpage
{\scriptsize\setlength{\tabcolsep}{3pt}
\begin{longtable}{@{}l r r r r r r l l r r r r@{}}
\caption{Cells of the crossed design. Runs: simulations with outcomes over planned simulations. Att.: attempts. Redr.: redraws after a fall. $\lS$ and ACI: mean and standard deviation over the runs of the cell (per unit of model time in M1, per stride in M2). $\alpha$: mean scaling exponent of the perturbation series (in), of the step durations and of the stride durations. Var.: state variability index, relative to the cell $d = 0$, $\sref$.}\label{tab:s5}\\
\toprule
Model & $d$ & $\sigma/\smax$ & Runs & Att. & Falls & Redr. & $\lS$, mean (SD) & ACI, mean (SD) & $\alpha_{\mathrm{in}}$ & $\alpha_{\mathrm{step}}$ & $\alpha_{\mathrm{stride}}$ & Var. \\
\midrule
\endfirsthead
\toprule
Model & $d$ & $\sigma/\smax$ & Runs & Att. & Falls & Redr. & $\lS$, mean (SD) & ACI, mean (SD) & $\alpha_{\mathrm{in}}$ & $\alpha_{\mathrm{step}}$ & $\alpha_{\mathrm{stride}}$ & Var. \\
\midrule
\endhead
\bottomrule
\endfoot
M1 & $-$0.30 & 0.2 & 10/10 & 10 & 0 & 0 & 0.472 (0.011) & 0.0055 (0.0097) & 0.26 & 0.14 & 0.14 & 0.35 \\
M1 & $-$0.20 & 0.2 & 10/10 & 10 & 0 & 0 & 0.485 (0.011) & 0.0053 (0.0090) & 0.29 & 0.17 & 0.18 & 0.34 \\
M1 & $-$0.10 & 0.2 & 10/10 & 10 & 0 & 0 & 0.491 (0.012) & 0.0137 (0.0072) & 0.42 & 0.28 & 0.29 & 0.33 \\
M1 & 0.00 & 0.2 & 20/20 & 20 & 0 & 0 & 0.485 (0.011) & 0.0141 (0.0066) & 0.53 & 0.39 & 0.44 & 0.32 \\
M1 & 0.10 & 0.2 & 10/10 & 10 & 0 & 0 & 0.485 (0.008) & 0.0214 (0.0086) & 0.56 & 0.45 & 0.48 & 0.31 \\
M1 & 0.20 & 0.2 & 10/10 & 10 & 0 & 0 & 0.489 (0.010) & 0.0279 (0.0096) & 0.69 & 0.58 & 0.62 & 0.29 \\
M1 & 0.30 & 0.2 & 10/10 & 10 & 0 & 0 & 0.496 (0.009) & 0.0332 (0.0087) & 0.82 & 0.73 & 0.76 & 0.27 \\
M1 & 0.40 & 0.2 & 10/10 & 10 & 0 & 0 & 0.496 (0.010) & 0.0422 (0.0087) & 0.86 & 0.79 & 0.83 & 0.24 \\
M1 & 0.45 & 0.2 & 10/10 & 10 & 0 & 0 & 0.489 (0.012) & 0.0443 (0.0081) & 0.94 & 0.88 & 0.89 & 0.23 \\
M1 & $-$0.30 & 0.4 & 10/10 & 10 & 0 & 0 & 0.542 (0.007) & 0.0064 (0.0082) & 0.26 & 0.14 & 0.15 & 0.71 \\
M1 & $-$0.20 & 0.4 & 10/10 & 10 & 0 & 0 & 0.545 (0.007) & 0.0038 (0.0053) & 0.31 & 0.19 & 0.20 & 0.69 \\
M1 & $-$0.10 & 0.4 & 10/10 & 10 & 0 & 0 & 0.550 (0.014) & 0.0184 (0.0079) & 0.40 & 0.27 & 0.29 & 0.68 \\
M1 & 0.00 & 0.4 & 20/20 & 20 & 0 & 0 & 0.554 (0.009) & 0.0186 (0.0098) & 0.48 & 0.34 & 0.37 & 0.66 \\
M1 & 0.10 & 0.4 & 10/10 & 10 & 0 & 0 & 0.559 (0.009) & 0.0202 (0.0063) & 0.58 & 0.46 & 0.52 & 0.62 \\
M1 & 0.20 & 0.4 & 10/10 & 10 & 0 & 0 & 0.559 (0.012) & 0.0306 (0.0054) & 0.70 & 0.59 & 0.63 & 0.59 \\
M1 & 0.30 & 0.4 & 10/10 & 10 & 0 & 0 & 0.560 (0.014) & 0.0335 (0.0071) & 0.72 & 0.64 & 0.67 & 0.54 \\
M1 & 0.40 & 0.4 & 10/10 & 10 & 0 & 0 & 0.567 (0.005) & 0.0451 (0.0094) & 0.86 & 0.79 & 0.81 & 0.49 \\
M1 & 0.45 & 0.4 & 10/10 & 10 & 0 & 0 & 0.554 (0.009) & 0.0466 (0.0050) & 0.90 & 0.83 & 0.86 & 0.46 \\
M1 & $-$0.30 & 0.6 & 10/10 & 10 & 0 & 0 & 0.579 (0.011) & 0.0041 (0.0106) & 0.23 & 0.16 & 0.17 & 1.07 \\
M1 & $-$0.20 & 0.6 & 10/10 & 10 & 0 & 0 & 0.587 (0.008) & 0.0092 (0.0082) & 0.31 & 0.19 & 0.21 & 1.06 \\
M1 & $-$0.10 & 0.6 & 10/10 & 10 & 0 & 0 & 0.584 (0.015) & 0.0131 (0.0081) & 0.38 & 0.26 & 0.26 & 1.04 \\
M1 & 0.00 & 0.6 & 20/20 & 20 & 0 & 0 & 0.588 (0.011) & 0.0183 (0.0101) & 0.51 & 0.38 & 0.41 & 1.00 \\
M1 & 0.10 & 0.6 & 10/10 & 10 & 0 & 0 & 0.597 (0.010) & 0.0217 (0.0089) & 0.58 & 0.46 & 0.49 & 0.94 \\
M1 & 0.20 & 0.6 & 10/10 & 10 & 0 & 0 & 0.596 (0.012) & 0.0278 (0.0071) & 0.65 & 0.55 & 0.59 & 0.90 \\
M1 & 0.30 & 0.6 & 10/10 & 10 & 0 & 0 & 0.602 (0.010) & 0.0334 (0.0074) & 0.82 & 0.72 & 0.77 & 0.83 \\
M1 & 0.40 & 0.6 & 10/10 & 10 & 0 & 0 & 0.599 (0.014) & 0.0423 (0.0073) & 0.82 & 0.74 & 0.75 & 0.75 \\
M1 & 0.45 & 0.6 & 10/10 & 10 & 0 & 0 & 0.588 (0.007) & 0.0473 (0.0088) & 0.93 & 0.86 & 0.90 & 0.67 \\
M1 & $-$0.30 & 0.8 & 10/10 & 10 & 0 & 0 & 0.602 (0.015) & 0.0107 (0.0101) & 0.25 & 0.18 & 0.20 & 1.48 \\
M1 & $-$0.20 & 0.8 & 10/10 & 10 & 0 & 0 & 0.603 (0.013) & 0.0142 (0.0069) & 0.34 & 0.21 & 0.24 & 1.42 \\
M1 & $-$0.10 & 0.8 & 10/10 & 10 & 0 & 0 & 0.612 (0.013) & 0.0134 (0.0100) & 0.41 & 0.27 & 0.31 & 1.42 \\
M1 & 0.00 & 0.8 & 20/20 & 20 & 0 & 0 & 0.616 (0.010) & 0.0156 (0.0081) & 0.50 & 0.35 & 0.39 & 1.36 \\
M1 & 0.10 & 0.8 & 10/10 & 10 & 0 & 0 & 0.627 (0.012) & 0.0211 (0.0062) & 0.56 & 0.45 & 0.48 & 1.29 \\
M1 & 0.20 & 0.8 & 10/10 & 10 & 0 & 0 & 0.623 (0.014) & 0.0266 (0.0108) & 0.65 & 0.53 & 0.58 & 1.23 \\
M1 & 0.30 & 0.8 & 10/10 & 10 & 0 & 0 & 0.623 (0.020) & 0.0313 (0.0053) & 0.80 & 0.70 & 0.75 & 1.12 \\
M1 & 0.40 & 0.8 & 10/10 & 10 & 0 & 0 & 0.615 (0.012) & 0.0449 (0.0102) & 0.89 & 0.81 & 0.85 & 0.98 \\
M1 & 0.45 & 0.8 & 10/10 & 10 & 0 & 0 & 0.616 (0.011) & 0.0440 (0.0071) & 0.89 & 0.82 & 0.85 & 0.94 \\
M1 & $-$0.30 & 1.0 & 10/10 & 10 & 0 & 0 & 0.625 (0.010) & 0.0112 (0.0105) & 0.25 & 0.23 & 0.25 & 1.93 \\
M1 & $-$0.20 & 1.0 & 10/10 & 10 & 0 & 0 & 0.621 (0.013) & 0.0181 (0.0090) & 0.32 & 0.28 & 0.32 & 1.91 \\
M1 & $-$0.10 & 1.0 & 10/10 & 10 & 0 & 0 & 0.625 (0.015) & 0.0139 (0.0086) & 0.39 & 0.27 & 0.29 & 1.80 \\
M1 & 0.00 & 1.0 & 20/20 & 20 & 0 & 0 & 0.635 (0.011) & 0.0202 (0.0099) & 0.48 & 0.36 & 0.38 & 1.75 \\
M1 & 0.10 & 1.0 & 10/10 & 10 & 0 & 0 & 0.643 (0.006) & 0.0236 (0.0063) & 0.62 & 0.49 & 0.54 & 1.68 \\
M1 & 0.20 & 1.0 & 10/10 & 10 & 0 & 0 & 0.647 (0.008) & 0.0237 (0.0091) & 0.67 & 0.54 & 0.58 & 1.56 \\
M1 & 0.30 & 1.0 & 10/10 & 10 & 0 & 0 & 0.648 (0.010) & 0.0347 (0.0074) & 0.78 & 0.67 & 0.71 & 1.42 \\
M1 & 0.40 & 1.0 & 10/10 & 10 & 0 & 0 & 0.636 (0.012) & 0.0396 (0.0070) & 0.85 & 0.76 & 0.76 & 1.28 \\
M1 & 0.45 & 1.0 & 10/10 & 10 & 0 & 0 & 0.637 (0.011) & 0.0484 (0.0047) & 0.97 & 0.90 & 0.94 & 1.18 \\
M2 & $-$0.30 & 0.2 & 10/10 & 10 & 0 & 0 & 2.989 (0.100) & 0.0056 (0.0072) & 0.26 & 0.24 & 0.21 & 0.26 \\
M2 & $-$0.20 & 0.2 & 10/10 & 10 & 0 & 0 & 2.795 (0.057) & 0.0050 (0.0103) & 0.30 & 0.29 & 0.24 & 0.28 \\
M2 & $-$0.10 & 0.2 & 10/10 & 10 & 0 & 0 & 2.640 (0.052) & 0.0099 (0.0142) & 0.42 & 0.39 & 0.31 & 0.30 \\
M2 & 0.00 & 0.2 & 20/20 & 20 & 0 & 0 & 2.525 (0.063) & 0.0173 (0.0105) & 0.53 & 0.46 & 0.40 & 0.34 \\
M2 & 0.10 & 0.2 & 10/10 & 10 & 0 & 0 & 2.417 (0.076) & 0.0187 (0.0103) & 0.56 & 0.49 & 0.42 & 0.37 \\
M2 & 0.20 & 0.2 & 10/10 & 10 & 0 & 0 & 2.262 (0.041) & 0.0275 (0.0175) & 0.68 & 0.59 & 0.51 & 0.46 \\
M2 & 0.30 & 0.2 & 10/10 & 10 & 0 & 0 & 2.071 (0.060) & 0.0414 (0.0134) & 0.83 & 0.72 & 0.63 & 0.56 \\
M2 & 0.40 & 0.2 & 10/10 & 10 & 0 & 0 & 1.975 (0.080) & 0.0413 (0.0137) & 0.85 & 0.74 & 0.67 & 0.62 \\
M2 & 0.45 & 0.2 & 10/10 & 10 & 0 & 0 & 1.826 (0.088) & 0.0519 (0.0152) & 0.94 & 0.80 & 0.70 & 0.71 \\
M2 & $-$0.30 & 0.4 & 10/10 & 10 & 0 & 0 & 3.151 (0.098) & 0.0057 (0.0087) & 0.25 & 0.27 & 0.24 & 0.53 \\
M2 & $-$0.20 & 0.4 & 10/10 & 10 & 0 & 0 & 3.052 (0.111) & 0.0111 (0.0081) & 0.31 & 0.30 & 0.25 & 0.54 \\
M2 & $-$0.10 & 0.4 & 10/10 & 10 & 0 & 0 & 2.933 (0.068) & 0.0192 (0.0141) & 0.40 & 0.37 & 0.31 & 0.59 \\
M2 & 0.00 & 0.4 & 20/20 & 20 & 0 & 0 & 2.791 (0.084) & 0.0212 (0.0117) & 0.48 & 0.43 & 0.38 & 0.66 \\
M2 & 0.10 & 0.4 & 10/10 & 10 & 0 & 0 & 2.627 (0.034) & 0.0282 (0.0114) & 0.56 & 0.50 & 0.45 & 0.75 \\
M2 & 0.20 & 0.4 & 10/10 & 10 & 0 & 0 & 2.511 (0.066) & 0.0417 (0.0177) & 0.69 & 0.60 & 0.53 & 0.90 \\
M2 & 0.30 & 0.4 & 10/10 & 10 & 0 & 0 & 2.343 (0.071) & 0.0438 (0.0138) & 0.73 & 0.63 & 0.56 & 1.02 \\
M2 & 0.40 & 0.4 & 10/10 & 10 & 0 & 0 & 2.141 (0.094) & 0.0580 (0.0175) & 0.87 & 0.76 & 0.67 & 1.25 \\
M2 & 0.45 & 0.4 & 10/10 & 10 & 0 & 0 & 2.057 (0.095) & 0.0639 (0.0163) & 0.90 & 0.76 & 0.66 & 1.30 \\
M2 & $-$0.30 & 0.6 & 10/10 & 10 & 0 & 0 & 3.248 (0.059) & 0.0168 (0.0079) & 0.24 & 0.25 & 0.20 & 0.79 \\
M2 & $-$0.20 & 0.6 & 10/10 & 10 & 0 & 0 & 3.153 (0.052) & 0.0154 (0.0080) & 0.31 & 0.29 & 0.26 & 0.82 \\
M2 & $-$0.10 & 0.6 & 10/10 & 10 & 0 & 0 & 3.040 (0.059) & 0.0172 (0.0071) & 0.38 & 0.37 & 0.31 & 0.88 \\
M2 & 0.00 & 0.6 & 20/20 & 20 & 0 & 0 & 2.894 (0.046) & 0.0314 (0.0141) & 0.51 & 0.47 & 0.41 & 1.00 \\
M2 & 0.10 & 0.6 & 10/10 & 10 & 0 & 0 & 2.783 (0.080) & 0.0345 (0.0120) & 0.60 & 0.53 & 0.47 & 1.13 \\
M2 & 0.20 & 0.6 & 10/10 & 10 & 0 & 0 & 2.622 (0.060) & 0.0381 (0.0149) & 0.64 & 0.57 & 0.50 & 1.26 \\
M2 & 0.30 & 0.6 & 10/10 & 10 & 0 & 0 & 2.444 (0.082) & 0.0604 (0.0136) & 0.82 & 0.70 & 0.65 & 1.59 \\
M2 & 0.40 & 0.6 & 10/10 & 10 & 0 & 0 & 2.309 (0.079) & 0.0643 (0.0141) & 0.81 & 0.72 & 0.62 & 1.80 \\
M2 & 0.45 & 0.6 & 10/10 & 10 & 0 & 0 & 2.088 (0.104) & 0.0699 (0.0182) & 0.93 & 0.79 & 0.72 & 2.22 \\
M2 & $-$0.30 & 0.8 & 10/10 & 10 & 0 & 0 & 3.349 (0.087) & 0.0280 (0.0107) & 0.26 & 0.30 & 0.27 & 1.05 \\
M2 & $-$0.20 & 0.8 & 10/10 & 10 & 0 & 0 & 3.217 (0.072) & 0.0144 (0.0154) & 0.34 & 0.32 & 0.29 & 1.12 \\
M2 & $-$0.10 & 0.8 & 10/10 & 10 & 0 & 0 & 3.157 (0.064) & 0.0294 (0.0123) & 0.41 & 0.38 & 0.34 & 1.19 \\
M2 & 0.00 & 0.8 & 20/20 & 20 & 0 & 0 & 3.048 (0.102) & 0.0300 (0.0142) & 0.49 & 0.47 & 0.41 & 1.30 \\
M2 & 0.10 & 0.8 & 10/10 & 10 & 0 & 0 & 2.867 (0.051) & 0.0362 (0.0149) & 0.57 & 0.54 & 0.48 & 1.48 \\
M2 & 0.20 & 0.8 & 10/10 & 10 & 0 & 0 & 2.753 (0.063) & 0.0409 (0.0147) & 0.64 & 0.58 & 0.52 & 1.74 \\
M2 & 0.30 & 0.8 & 10/10 & 10 & 0 & 0 & 2.549 (0.067) & 0.0682 (0.0131) & 0.80 & 0.74 & 0.67 & 2.10 \\
M2 & 0.40 & 0.8 & 10/10 & 10 & 0 & 0 & 2.310 (0.078) & 0.0777 (0.0176) & 0.88 & 0.81 & 0.74 & 2.59 \\
M2 & 0.45 & 0.8 & 10/10 & 10 & 0 & 0 & 2.257 (0.110) & 0.0733 (0.0139) & 0.90 & 0.79 & 0.71 & 2.73 \\
M2 & $-$0.30 & 1.0 & 10/10 & 10 & 0 & 0 & 3.449 (0.053) & 0.0332 (0.0104) & 0.24 & 0.30 & 0.27 & 1.32 \\
M2 & $-$0.20 & 1.0 & 10/10 & 10 & 0 & 0 & 3.328 (0.066) & 0.0352 (0.0085) & 0.32 & 0.35 & 0.32 & 1.39 \\
M2 & $-$0.10 & 1.0 & 10/10 & 10 & 0 & 0 & 3.222 (0.095) & 0.0328 (0.0124) & 0.39 & 0.39 & 0.33 & 1.51 \\
M2 & 0.00 & 1.0 & 20/20 & 20 & 0 & 0 & 3.099 (0.078) & 0.0344 (0.0128) & 0.48 & 0.47 & 0.42 & 1.64 \\
M2 & 0.10 & 1.0 & 10/10 & 10 & 0 & 0 & 2.931 (0.083) & 0.0420 (0.0124) & 0.61 & 0.57 & 0.52 & 1.93 \\
M2 & 0.20 & 1.0 & 10/10 & 10 & 0 & 0 & 2.782 (0.075) & 0.0481 (0.0098) & 0.65 & 0.61 & 0.55 & 2.22 \\
M2 & 0.30 & 1.0 & 10/10 & 10 & 0 & 0 & 2.615 (0.054) & 0.0672 (0.0083) & 0.79 & 0.71 & 0.64 & 2.67 \\
M2 & 0.40 & 1.0 & 10/10 & 10 & 0 & 0 & 2.411 (0.101) & 0.0746 (0.0137) & 0.85 & 0.80 & 0.67 & 3.16 \\
M2 & 0.45 & 1.0 & 10/10 & 10 & 0 & 0 & 2.303 (0.104) & 0.0847 (0.0211) & 0.96 & 0.86 & 0.80 & 3.56 \\
\end{longtable}}

{\scriptsize
\begin{longtable}{@{}l l l l l l@{}}
\caption{Secondary analysis: the criteria of H1 in each slice at fixed $d$ and those of H2 in each slice at fixed $\sigma$ (rad in M1, leg length in M2). Each range ratio divides the range within the slice by the denominator of the primary ratio. Intervals: 95\% percentile bootstrap intervals.}\label{tab:s6}\\
\toprule
Model & Criterion & Slice & Statistic & Estimate [95\% interval] & Met \\
\midrule
\endfirsthead
\toprule
Model & Criterion & Slice & Statistic & Estimate [95\% interval] & Met \\
\midrule
\endhead
\bottomrule
\endfoot
M1 & H1a & $d = -0.30$ & $\eta^2$ of $\lS$ on $\sigma$ & 0.962 [0.952, 0.981] & yes \\
M1 & H1a & $d = -0.30$ & $\rho$ of $\lS$ with $\sigma$ & 0.959 [0.937, 0.977] & yes \\
M1 & H1a & $d = -0.20$ & $\eta^2$ of $\lS$ on $\sigma$ & 0.958 [0.945, 0.979] & yes \\
M1 & H1a & $d = -0.20$ & $\rho$ of $\lS$ with $\sigma$ & 0.944 [0.910, 0.970] & yes \\
M1 & H1a & $d = -0.10$ & $\eta^2$ of $\lS$ on $\sigma$ & 0.931 [0.915, 0.960] & yes \\
M1 & H1a & $d = -0.10$ & $\rho$ of $\lS$ with $\sigma$ & 0.935 [0.894, 0.966] & yes \\
M1 & H1a & $d = 0.00$ & $\eta^2$ of $\lS$ on $\sigma$ & 0.964 [0.956, 0.974] & yes \\
M1 & H1a & $d = 0.00$ & $\rho$ of $\lS$ with $\sigma$ & 0.967 [0.956, 0.977] & yes \\
M1 & H1a & $d = 0.10$ & $\eta^2$ of $\lS$ on $\sigma$ & 0.977 [0.968, 0.989] & yes \\
M1 & H1a & $d = 0.10$ & $\rho$ of $\lS$ with $\sigma$ & 0.968 [0.948, 0.981] & yes \\
M1 & H1a & $d = 0.20$ & $\eta^2$ of $\lS$ on $\sigma$ & 0.965 [0.954, 0.981] & yes \\
M1 & H1a & $d = 0.20$ & $\rho$ of $\lS$ with $\sigma$ & 0.971 [0.958, 0.981] & yes \\
M1 & H1a & $d = 0.30$ & $\eta^2$ of $\lS$ on $\sigma$ & 0.947 [0.930, 0.972] & yes \\
M1 & H1a & $d = 0.30$ & $\rho$ of $\lS$ with $\sigma$ & 0.951 [0.927, 0.972] & yes \\
M1 & H1a & $d = 0.40$ & $\eta^2$ of $\lS$ on $\sigma$ & 0.958 [0.946, 0.976] & yes \\
M1 & H1a & $d = 0.40$ & $\rho$ of $\lS$ with $\sigma$ & 0.945 [0.914, 0.969] & yes \\
M1 & H1a & $d = 0.45$ & $\eta^2$ of $\lS$ on $\sigma$ & 0.967 [0.952, 0.985] & yes \\
M1 & H1a & $d = 0.45$ & $\rho$ of $\lS$ with $\sigma$ & 0.966 [0.948, 0.981] & yes \\
M1 & H1b & $d = -0.30$ & range ratio of the ACI & 0.163 [0.107, 0.372] & yes \\
M1 & H1b & $d = -0.20$ & range ratio of the ACI & 0.333 [0.218, 0.510] & no \\
M1 & H1b & $d = -0.10$ & range ratio of the ACI & 0.123 [0.074, 0.325] & yes \\
M1 & H1b & $d = 0.00$ & range ratio of the ACI & 0.142 [0.079, 0.289] & yes \\
M1 & H1b & $d = 0.10$ & range ratio of the ACI & 0.079 [0.051, 0.250] & yes \\
M1 & H1b & $d = 0.20$ & range ratio of the ACI & 0.160 [0.087, 0.336] & yes \\
M1 & H1b & $d = 0.30$ & range ratio of the ACI & 0.080 [0.054, 0.246] & yes \\
M1 & H1b & $d = 0.40$ & range ratio of the ACI & 0.127 [0.071, 0.323] & yes \\
M1 & H1b & $d = 0.45$ & range ratio of the ACI & 0.101 [0.063, 0.255] & yes \\
M1 & H2a & $\sigma = 7.80 \times 10^{-4}$ & $\eta^2$ of the ACI on $d$ & 0.742 [0.695, 0.826] & yes \\
M1 & H2a & $\sigma = 7.80 \times 10^{-4}$ & $\rho$ of the ACI with $d$ & 0.846 [0.791, 0.890] & yes \\
M1 & H2a & $\sigma = 1.56 \times 10^{-3}$ & $\eta^2$ of the ACI on $d$ & 0.782 [0.739, 0.857] & yes \\
M1 & H2a & $\sigma = 1.56 \times 10^{-3}$ & $\rho$ of the ACI with $d$ & 0.857 [0.804, 0.904] & yes \\
M1 & H2a & $\sigma = 2.34 \times 10^{-3}$ & $\eta^2$ of the ACI on $d$ & 0.719 [0.667, 0.806] & yes \\
M1 & H2a & $\sigma = 2.34 \times 10^{-3}$ & $\rho$ of the ACI with $d$ & 0.841 [0.787, 0.885] & yes \\
M1 & H2a & $\sigma = 3.12 \times 10^{-3}$ & $\eta^2$ of the ACI on $d$ & 0.687 [0.627, 0.790] & yes \\
M1 & H2a & $\sigma = 3.12 \times 10^{-3}$ & $\rho$ of the ACI with $d$ & 0.769 [0.692, 0.835] & yes \\
M1 & H2a & $\sigma = 3.90 \times 10^{-3}$ & $\eta^2$ of the ACI on $d$ & 0.661 [0.607, 0.768] & yes \\
M1 & H2a & $\sigma = 3.90 \times 10^{-3}$ & $\rho$ of the ACI with $d$ & 0.748 [0.672, 0.819] & yes \\
M1 & H2b & $\sigma = 7.80 \times 10^{-4}$ & range ratio of $\lS$ & 0.158 [0.122, 0.231] & yes \\
M1 & H2b & $\sigma = 1.56 \times 10^{-3}$ & range ratio of $\lS$ & 0.170 [0.140, 0.211] & yes \\
M1 & H2b & $\sigma = 2.34 \times 10^{-3}$ & range ratio of $\lS$ & 0.152 [0.110, 0.218] & yes \\
M1 & H2b & $\sigma = 3.12 \times 10^{-3}$ & range ratio of $\lS$ & 0.164 [0.130, 0.259] & yes \\
M1 & H2b & $\sigma = 3.90 \times 10^{-3}$ & range ratio of $\lS$ & 0.181 [0.149, 0.253] & yes \\
M2 & H1a & $d = -0.30$ & $\eta^2$ of $\lS$ on $\sigma$ & 0.808 [0.745, 0.896] & yes \\
M2 & H1a & $d = -0.30$ & $\rho$ of $\lS$ with $\sigma$ & 0.907 [0.852, 0.951] & yes \\
M2 & H1a & $d = -0.20$ & $\eta^2$ of $\lS$ on $\sigma$ & 0.868 [0.820, 0.938] & yes \\
M2 & H1a & $d = -0.20$ & $\rho$ of $\lS$ with $\sigma$ & 0.901 [0.806, 0.960] & yes \\
M2 & H1a & $d = -0.10$ & $\eta^2$ of $\lS$ on $\sigma$ & 0.907 [0.883, 0.946] & yes \\
M2 & H1a & $d = -0.10$ & $\rho$ of $\lS$ with $\sigma$ & 0.925 [0.887, 0.960] & yes \\
M2 & H1a & $d = 0.00$ & $\eta^2$ of $\lS$ on $\sigma$ & 0.882 [0.859, 0.913] & yes \\
M2 & H1a & $d = 0.00$ & $\rho$ of $\lS$ with $\sigma$ & 0.918 [0.892, 0.941] & yes \\
M2 & H1a & $d = 0.10$ & $\eta^2$ of $\lS$ on $\sigma$ & 0.892 [0.871, 0.934] & yes \\
M2 & H1a & $d = 0.10$ & $\rho$ of $\lS$ with $\sigma$ & 0.916 [0.865, 0.958] & yes \\
M2 & H1a & $d = 0.20$ & $\eta^2$ of $\lS$ on $\sigma$ & 0.912 [0.893, 0.945] & yes \\
M2 & H1a & $d = 0.20$ & $\rho$ of $\lS$ with $\sigma$ & 0.920 [0.880, 0.953] & yes \\
M2 & H1a & $d = 0.30$ & $\eta^2$ of $\lS$ on $\sigma$ & 0.899 [0.877, 0.939] & yes \\
M2 & H1a & $d = 0.30$ & $\rho$ of $\lS$ with $\sigma$ & 0.916 [0.880, 0.948] & yes \\
M2 & H1a & $d = 0.40$ & $\eta^2$ of $\lS$ on $\sigma$ & 0.777 [0.719, 0.867] & yes \\
M2 & H1a & $d = 0.40$ & $\rho$ of $\lS$ with $\sigma$ & 0.836 [0.770, 0.896] & yes \\
M2 & H1a & $d = 0.45$ & $\eta^2$ of $\lS$ on $\sigma$ & 0.759 [0.705, 0.858] & yes \\
M2 & H1a & $d = 0.45$ & $\rho$ of $\lS$ with $\sigma$ & 0.879 [0.820, 0.929] & yes \\
M2 & H1b & $d = -0.30$ & range ratio of the ACI & 0.506 [0.377, 0.677] & no \\
M2 & H1b & $d = -0.20$ & range ratio of the ACI & 0.554 [0.386, 0.712] & no \\
M2 & H1b & $d = -0.10$ & range ratio of the ACI & 0.420 [0.257, 0.623] & no \\
M2 & H1b & $d = 0.00$ & range ratio of the ACI & 0.315 [0.206, 0.461] & no \\
M2 & H1b & $d = 0.10$ & range ratio of the ACI & 0.427 [0.270, 0.609] & no \\
M2 & H1b & $d = 0.20$ & range ratio of the ACI & 0.379 [0.197, 0.617] & no \\
M2 & H1b & $d = 0.30$ & range ratio of the ACI & 0.494 [0.374, 0.728] & no \\
M2 & H1b & $d = 0.40$ & range ratio of the ACI & 0.669 [0.472, 0.919] & no \\
M2 & H1b & $d = 0.45$ & range ratio of the ACI & 0.602 [0.345, 0.892] & no \\
M2 & H2a & $\sigma = 6.00 \times 10^{-5}$ & $\eta^2$ of the ACI on $d$ & 0.625 [0.566, 0.737] & yes \\
M2 & H2a & $\sigma = 6.00 \times 10^{-5}$ & $\rho$ of the ACI with $d$ & 0.767 [0.695, 0.828] & yes \\
M2 & H2a & $\sigma = 1.20 \times 10^{-4}$ & $\eta^2$ of the ACI on $d$ & 0.677 [0.624, 0.783] & yes \\
M2 & H2a & $\sigma = 1.20 \times 10^{-4}$ & $\rho$ of the ACI with $d$ & 0.814 [0.765, 0.861] & yes \\
M2 & H2a & $\sigma = 1.80 \times 10^{-4}$ & $\eta^2$ of the ACI on $d$ & 0.710 [0.647, 0.819] & yes \\
M2 & H2a & $\sigma = 1.80 \times 10^{-4}$ & $\rho$ of the ACI with $d$ & 0.831 [0.778, 0.879] & yes \\
M2 & H2a & $\sigma = 2.40 \times 10^{-4}$ & $\eta^2$ of the ACI on $d$ & 0.705 [0.663, 0.795] & yes \\
M2 & H2a & $\sigma = 2.40 \times 10^{-4}$ & $\rho$ of the ACI with $d$ & 0.749 [0.685, 0.810] & yes \\
M2 & H2a & $\sigma = 3.00 \times 10^{-4}$ & $\eta^2$ of the ACI on $d$ & 0.700 [0.642, 0.798] & yes \\
M2 & H2a & $\sigma = 3.00 \times 10^{-4}$ & $\rho$ of the ACI with $d$ & 0.758 [0.689, 0.823] & yes \\
M2 & H2b & $\sigma = 6.00 \times 10^{-5}$ & range ratio of $\lS$ & 2.024 [1.833, 2.241] & no \\
M2 & H2b & $\sigma = 1.20 \times 10^{-4}$ & range ratio of $\lS$ & 1.902 [1.716, 2.118] & no \\
M2 & H2b & $\sigma = 1.80 \times 10^{-4}$ & range ratio of $\lS$ & 2.017 [1.832, 2.227] & no \\
M2 & H2b & $\sigma = 2.40 \times 10^{-4}$ & range ratio of $\lS$ & 1.901 [1.721, 2.100] & no \\
M2 & H2b & $\sigma = 3.00 \times 10^{-4}$ & range ratio of $\lS$ & 1.994 [1.808, 2.182] & no \\
\end{longtable}}

\begin{table}[p]
\centering
\caption{Analyses of the registered plan (sections in parentheses) and the place where each is reported. State of the work on 13 November 2026. The sensitivity analyses of the plan are all exploratory.}
\label{tab:s7}
\small
\begin{tabular}{@{}>{\raggedright\arraybackslash}p{0.47\linewidth}>{\raggedright\arraybackslash}p{0.47\linewidth}@{}}
\toprule
Registered analysis & Report \\
\midrule
Replication of M1 and of M2 (4.2, 6.1) & this article, Appendices~\ref{app:m1} and \ref{app:m2} \\
H1 and H2 in M1 and M2, with the analysis by slice (6.3, 6.4) & this article, Sections~\ref{sec:res_h1} and \ref{sec:res_h2}; Tables~\ref{tab:decisions} and \ref{tab:s6} \\
Human stride series, S1 and S2, with their three sensitivity analyses (4.8, 6.7) & this article, Section~\ref{sec:res_human} \\
Human stride series, exploratory analyses (6.7): metronome series against their surrogates & this article, Section~\ref{sec:res_human} \\
Human stride series, exploratory analyses (6.7): $\lS$ for all series & this article, Table~\ref{tab:s3} \\
Human stride series, exploratory analyses (6.7): ACI against the parameters of a synchronization model fitted to each series & computed; not reported in this article \\
Removal of the cells flagged for falls (4.4, 9) & this article: no cell was flagged (Table~\ref{tab:s5}) \\
Estimation settings of M2 in its replication runs (5.1, 9) & this article, Appendix~\ref{app:m2} \\
Fixed set of pairs in the estimator (5.1, 9) & this article, for the crossed design (Table~\ref{tab:s9}); not computed for the human stride series \\
\addlinespace
H1 and H2 in two further models, M3 and M4 (4.1, 4.7) & not run; second article \\
H3, closed loop with an endogenous controller (4.6, 6.5) & not run; second article \\
AR family of perturbations (4.9, 6.8) & not run \\
Agreement of the ACI with the analytic predictor (5.3, 6.9) & computed for the human stride series and for the crossed design; not reported in this article (second article) \\
Series-length requirement (7) & not run \\
H4 on recordings of daily-life walking (8) & not run; outside this article \\
\addlinespace
Other slope method of each model (9) & not computed for the crossed design \\
Uniform white noise at $d = 0$ (9) & not computed \\
Neighbor exclusion, number of neighbors, delay embedding (9) & not computed for the crossed design \\
Map of the fitting window over the divergence curve (9) & not computed \\
Other injection point (9) & not computed \\
\bottomrule
\end{tabular}
\end{table}

\begin{table}[tp]
\centering
\caption{Simulations with the human stride series. A series that made a model fall was excluded with its pair; a surrogate that fell was replaced by the next permutation.}
\label{tab:s8}
\small
\begin{tabular}{@{}l l l r r r r r@{}}
\toprule
Model & Condition & Kind & Runs & Falls & Without outcomes & Excluded with their pair & Kept \\
\midrule
M1 & free & original & 30 & 0 & 0 & 2 & 28 \\
M1 & free & surrogate & 280 & 0 & 0 & 0 & 280 \\
M1 & metronome & original & 30 & 2 & 2 & 2 & 28 \\
M1 & metronome & surrogate & 283 & 3 & 3 & 0 & 280 \\
M2 & free & original & 30 & 0 & 0 & 2 & 28 \\
M2 & free & surrogate & 280 & 0 & 0 & 0 & 280 \\
M2 & metronome & original & 30 & 2 & 2 & 2 & 28 \\
M2 & metronome & surrogate & 280 & 0 & 0 & 0 & 280 \\
\bottomrule
\end{tabular}
\end{table}

\begin{table}[tp]
\centering
\caption{Sensitivity analysis with a fixed set of pairs, crossed design. Registered criteria of H1 and H2, with the shrinking set of pairs of the primary analysis and with the fixed set. Estimates with their 95\% percentile intervals (bootstrap within cells). Met: the point estimate meets the threshold.}
\label{tab:s9}
\footnotesize
\setlength{\tabcolsep}{3pt}
\begin{tabular}{@{}l l l >{\raggedright\arraybackslash}p{0.14\linewidth} l l l l@{}}
\toprule
Model & Criterion & Statistic & Threshold & Shrinking set & Met & Fixed set & Met \\
\midrule
M1 & H1a & $\eta^2$ & $\geq$ 0.50 & 0.964 [0.956, 0.974] & yes & 0.964 [0.956, 0.974] & yes \\
M1 & H1a & Spearman $\rho$ & $>$ 0 and interval above 0 & 0.967 [0.956, 0.977] & yes & 0.966 [0.955, 0.976] & yes \\
M1 & H1b & range ratio & $\leq$ 0.25 & 0.142 [0.079, 0.289] & yes & 0.144 [0.077, 0.285] & yes \\
M1 & H2a & $\eta^2$ & $\geq$ 0.50 & 0.719 [0.667, 0.806] & yes & 0.722 [0.669, 0.810] & yes \\
M1 & H2a & Spearman $\rho$ & $>$ 0 and interval above 0 & 0.841 [0.787, 0.885] & yes & 0.841 [0.786, 0.886] & yes \\
M1 & H2b & range ratio & $\leq$ 0.25 & 0.152 [0.110, 0.218] & yes & 0.148 [0.107, 0.216] & yes \\
\addlinespace
M2 & H1a & $\eta^2$ & $\geq$ 0.50 & 0.882 [0.859, 0.913] & yes & 0.886 [0.865, 0.916] & yes \\
M2 & H1a & Spearman $\rho$ & $>$ 0 and interval above 0 & 0.918 [0.892, 0.941] & yes & 0.924 [0.901, 0.944] & yes \\
M2 & H1b & range ratio & $\leq$ 0.25 & 0.315 [0.206, 0.461] & no & 0.312 [0.197, 0.450] & no \\
M2 & H2a & $\eta^2$ & $\geq$ 0.50 & 0.710 [0.647, 0.819] & yes & 0.714 [0.656, 0.817] & yes \\
M2 & H2a & Spearman $\rho$ & $>$ 0 and interval above 0 & 0.831 [0.778, 0.879] & yes & 0.828 [0.777, 0.873] & yes \\
M2 & H2b & range ratio & $\leq$ 0.25 & 2.017 [1.832, 2.227] & no & 1.996 [1.807, 2.217] & no \\
\addlinespace
\bottomrule
\end{tabular}
\end{table}

\end{document}